%% file: main.tex
\documentclass{article}

\usepackage[final]{corl_2020} % Uncomment for the camera-ready ``final'' version.

\usepackage{amsfonts}
\usepackage{array}
\usepackage{graphicx}
\usepackage{multirow}
\usepackage{subcaption}
\usepackage{url}
\usepackage{wrapfig}
\usepackage{xcolor}
\usepackage{booktabs}
\usepackage{breakurl}

\newcommand{\platform}{SMARTS}

\title{\platform{}: Scalable Multi-Agent Reinforcement Learning Training School for Autonomous Driving}

\author{\hspace{-25pt}
    Ming Zhou\thanks{Equal contribution. $\dag$ Work done as an intern at Huawei. $\ddag$ Corresponding author: \href{jun.luo1@huawei.com}{jun.luo1@huawei.com}.} \textsuperscript{ \rm $\dag$1,2}, 
    Jun Luo\textsuperscript{*$\ddag$\rm 1},
    Julian Villella\textsuperscript{*\rm 1},
    Yaodong Yang\textsuperscript{*\rm 1,3},
    David Rusu\textsuperscript{\rm 1},
    Jiayu Miao\textsuperscript{\rm $\dag$1,2} \\
    \normalsize \hspace{-20pt}\textbf{
    Weinan Zhang\textsuperscript{\rm 2},
    Montgomery Alban\textsuperscript{\rm 1},
    Iman Fadakar\textsuperscript{\rm 1},
    Zheng Chen\textsuperscript{\rm 1},
    Aurora C. Huang\textsuperscript{\rm 1} %Aurora Chongxi Huang
    } \\
    \normalsize \hspace{-20pt}\textbf{
    Ying Wen\textsuperscript{\rm $\dag$1,2},
    Kimia Hassanzadeh\textsuperscript{\rm 1},
    Daniel Graves\textsuperscript{\rm 1},
    Dong Chen\textsuperscript{\rm 1},
    Zhengbang Zhu\textsuperscript{\rm $\dag$1,2}
    }\\
    \normalsize \hspace{-20pt}\textbf{
    Nhat Nguyen\textsuperscript{\rm 1},
    Mohamed Elsayed\textsuperscript{\rm $\dag$1},
    Kun Shao\textsuperscript{\rm 1},
    Sanjeevan Ahilan\textsuperscript{\rm $\dag$1},
    Baokuan Zhang\textsuperscript{\rm 1},
    Jiannan Wu\textsuperscript{\rm 1}}
    \\
    \normalsize \hspace{-20pt}\textbf{
    Zhengang Fu\textsuperscript{\rm 1},
    Kasra Rezaee\textsuperscript{\rm 1}, 
    Peyman Yadmellat\textsuperscript{\rm 1},
    Mohsen Rohani\textsuperscript{\rm 1},
    Nicolas Perez Nieves \textsuperscript{\rm 1}
    }\\
    \normalsize \hspace{-20pt}\textbf{
    Yihan Ni\textsuperscript{\rm $\dag$1},
    Seyedershad Banijamali\textsuperscript{\rm 1},
    Alexander I.  Cowen-Rivers\textsuperscript{\rm 1}, %Alexander Cowen Rivers
    Zheng Tian\textsuperscript{\rm $\dag$1,3},
    Daniel Palenicek\textsuperscript{\rm $\dag$1}
    }\\
    \normalsize \hspace{-20pt}\textbf{
    Haitham bou Ammar\textsuperscript{\rm 1,3},
    Hongbo Zhang\textsuperscript{\rm 1}, 
    Wulong Liu\textsuperscript{\rm 1}, 
    Jianye Hao\textsuperscript{\rm 1}, 
    Jun Wang\textsuperscript{\rm 1,3}
    }\\
\\
\hspace{-13pt}
\textsuperscript{\rm 1}Noah's Ark Lab, Huawei Technologies, \textsuperscript{\rm 2}Shanghai Jiao Tong University, \textsuperscript{\rm 3}University College London\\
\setcounter{footnote}{0}

}

\begin{document}
\maketitle

%===============================================================================

\begin{abstract}
    Multi-agent interaction is a fundamental aspect of autonomous driving in the real world. Despite more than a decade of research and development, the problem of how to competently interact with diverse road users in diverse scenarios remains largely unsolved. Learning methods have much to offer towards solving this problem. But they require a realistic multi-agent simulator that generates diverse and competent driving interactions. To meet this need, we develop a dedicated simulation platform called \platform{} (\textbf{S}calable \textbf{M}ulti-\textbf{A}gent \textbf{R}L \textbf{T}raining \textbf{S}chool). SMARTS supports the training, accumulation, and use of diverse behavior models of road users. These are in turn used to create increasingly more realistic and diverse interactions that enable deeper and broader research on multi-agent interaction. In this paper, we describe the design goals of \platform{}, explain its basic architecture and its key features, and illustrate its use through concrete multi-agent experiments on interactive scenarios. We open-source the \platform{} platform and the associated benchmark tasks and evaluation metrics to encourage and empower research on multi-agent learning for autonomous driving. Our code is available at \href{https://github.com/huawei-noah/SMARTS}{https://github.com/huawei-noah/SMARTS}. 
\end{abstract}

\keywords{autonomous driving, simulation, multi-agent, reinforcement learning}

\input{introduction}

\input{platform}

\input{benchmarking}

\input{experiments}

\input{conclusion}

%===============================================================================

% The maximum paper length is 8 pages excluding references and acknowledgements, and 10 pages including references and acknowledgements

%\clearpage
% The acknowledgments are automatically included only in the final version of the paper.
\acknowledgments{The authors would like to thank the anonymous reviewers for their helpful comments and thank the many Huawei colleagues, SJTU students, competition participants in U.K. and China, and other early users of \platform{} for their valuable feedback. Weinan Zhang's contribution to this work is partly supported by ``New Generation of AI 2030'' Major Project (2018AAA0100900), National Natural Science Foundation of China (61632017) and Huawei Innovation Research Program.}

%If a paper is accepted, the final camera-ready version will (and probably should) include acknowledgments. All acknowledgments go at the end of the paper, including thanks to reviewers who gave useful comments, to colleagues who contributed to the ideas, and to funding agencies and corporate sponsors that provided financial support.
%===============================================================================

\small{
\bibliography{references}  % .bib
}
\include{appendices}

\end{document}

%% file: introduction.tex
%===============================================================================

\section{Introduction}

Sixteen years have passed since the DAPRA Grand Challenge \cite{buehler20072005}. More than two millions of autonomous miles have been driven on public roads by Waymo's cars alone \citep{waymo2020safety}. In spite of such remarkable achievement, fundamental challenges remain underexplored. One is the weather: to date, most deployment of autonomous driving (AD) has been in fair weather \citep{badue2020self}. Another challenge is realistic interaction with diverse road users. Current mainstream level-4 AD solutions tend to limit interaction rather than embrace it: when encountering complexly interactive scenarios, the autonomous car tends to slow down and wait rather than acting proactively to find another way through. While this strategy is practical enough to put autonomous cars on many streets without getting into accidents where they are at fault, its conservativeness also creates headaches to other road users and compromises rider experience. In California in 2018, 57\% of autonomous car crashes are rear endings and 29\% are sideswipes---all by other cars and thus attributable to the conservativeness of the autonomous car \cite{stewart}. Consider Waymo, for example, whose leadership position and longer history also gave us more publicly available data points. Issues with interaction surfaced from their trials in Arizona and California, including difficulties with unprotected left turns, merging into highway traffic, and road users who do not follow traffic regulations \cite{WaymoB72}. It was reported that Waymo vehicles tend to form a trail of other cars behind them on small roads, causing other drivers to illegally pull around them \cite{Someofth6}. In addition, Waymo vehicles often brake excessively compared to human standard, sometimes making riders and passengers carsick \cite{WaymoRid1}. Finally, the pick-up and drop-off locations are often limited, adding significant amount of total travel time \cite{WaymoB80}.

To face up to the interaction challenge in AD, we take the view that driving in shared public space with diverse road users is fundamentally a \textit{multi-agent problem} that requires \textit{multi-agent learning} as a key part of the solution. To illustrate this perspective, we propose---with a nod to \citep{sae2018levels}---the ``multi-agent learning levels'', or ``M-levels'' for short. 
\textbf{M0} agents are designed to follow specific rules and stick to them regardless of how the environment dynamics may have shifted. 
After driving many times through a dense intersection, for example, the agents have no improvements except through tweaks by human engineers. 
\textbf{M1} agents can learn to adapt from its online experiences and such a learning process is expected to change their behaviors for future runs. 
\textbf{M2} agents not only learn to adapt, but also learn to model other road users. However, there is still no direct information exchange among the learning agents during such a decentralized learning process. 
\textbf{M3} further requires information exchange among the agents during training so as to coordinate their learning, but at execution time there is neither centralized control nor direct information exchange.
At \textbf{M4}, agents are required to coordinate their learning at the local group level (e.g. at an intersection) in a way such that the Nash equilibrium or other equilibrium variant is ensured at execution time.
At \textbf{M5}, agents start to focus on solving how their local actions in the scenario (e.g. ``how I change into the other lane here and now'') may impact global welfare (e.g. ``may decide whether there is going to be a congestion on the highway five minutes later'') so as to minimise \emph{the price of anarchy} \cite{roughgarden2005selfish}.

\begin{table}[t!]
\footnotesize  
\begin{center}
\small
\begin{tabular}{|>{\centering}m{0.8cm} | m{3.3cm} | m{8.4cm} |}
    \hline
    \textbf{Level} & \textbf{Description} & \textbf{Possible MARL Approach illustrated with Double Merge} \\
    \hline
    \textbf{M0} & Rule-based planning and control without learning & N/A \\
    \hline
    \textbf{M1} & Single-agent learning without coordination with other learning agents & An agent could learn to implicitly anticipate how other agents will react to its own actions but the learned solution will likely suffer from non-stationarity and lack generalizability \cite{kendall2019learning}. \\
    \hline
    \textbf{M2} & Decentralized multi-agent learning with opponent modeling & MARL to model other agents, e.g. ``how likely are they to yield to me if I start changing to their lane given how they have been driving in the past few seconds?'' \cite{mobileye2016} \cite{schwarting2019} \\
    \hline
    \textbf{M3} & Coordinated multi-agent learning and independent execution & Coordinated learning of what to expect of each other even if there is no explicit coordination during execution, e.g. ``some of them will give me the gap.''  \cite{wang2020r} \\
    \hline
    \textbf{M4} & Local (Nash) equilibrium oriented multi-agent learning & Learn as a group towards a certain equilibrium for the double merge such that the cars from the two sides will weave through each other without much trouble. \cite{AIMAuton91:online,zhang2019bi} \\
    \hline
    \textbf{M5} & Social welfare oriented multi-agent learning & Learn broader repercussions of our actions, e.g. ``if I force to the right lane now, I will cause a congestion on the highway, due to the fast heavy traffic coming down to this double merge from there.''\\
    \hline
\end{tabular}
\end{center}
\caption{Levels of multi-agent learning in autonomous driving.}
\label{table:m-levels}
\vspace{-3mm}
\end{table}

\begin{wrapfigure}{r}{0.4\textwidth}
  \vspace{-4mm}
    \centering
    \includegraphics[width=1.0\linewidth]{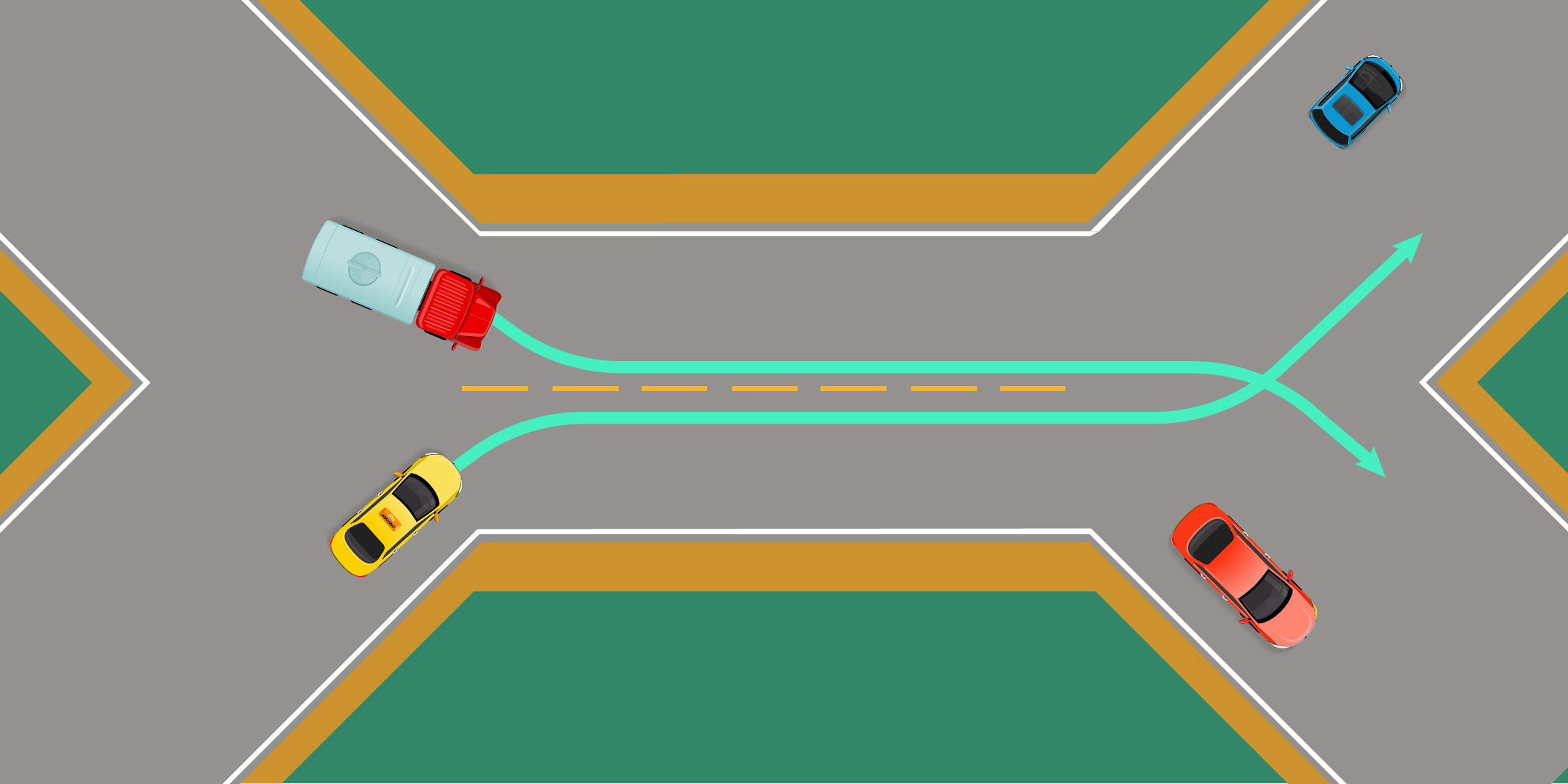}
    \captionof{figure}{The double merge scenario.}
    \label{fig:double-merge}
  \vspace{-3mm}
\end{wrapfigure}
Consider the so-called ``double merge'' (Figure \ref{fig:double-merge}) studied in \citet{mobileye2016}. There are vehicles coming into a shared road section from one side and aiming to exit this section on the opposite side. In the process, they need to change lane by weaving through cars that may also be changing lane in the opposite direction. Failing that, they will be forced to either go out on the wrong road or wait somewhere before the road splits until a usable gap emerges.\footnote{Such predicament does happen to autonomous cars. See a case in which an vehicle was forced to wait right before the split: \href{https://youtu.be/HjtiiGCe1pE}{https://youtu.be/HjtiiGCe1pE}, and a related case: \href{https://youtu.be/spw176TZ7-8?t=90}{https://youtu.be/spw176TZ7-8?t=90}.} Double merge is a sequential decision making setting that involves complex interaction among multiple agents. Questions one can ask about it include: Should a car drive further or wait here when looking for a gap? Should it force a lane change even when there is no adequate gap? Will the other cars give the gap? Is this other car coming over to my lane so that I can trade places with it? Furthermore, handwritten rules are unlikely to scale up to the full complexity of interactive scenarios such as the double merge \cite{mobileye2016}. We think learning with consideration of interaction is necessary and multi-agent reinforcement learning (MARL) is especially promising. %\yaodong{it is unclear to me why we need multi-agent here? worth putting the simple matrix game of rush-yield here to illustrate?}. \JL: we have been using multi-agent both "formally" and "informally". My personal sense is that for a situation to be multi-agent that requires multi-agent treatment, the matrix game sort of formulation could be both unnecessary and unhelpful. This does not mean multi-agent formalisms won't be helpful, but that they are probably all limited and should be expanded as research moves forward. 
To show how MARL could be specifically relevant to autonomous driving, we have in Table \ref{table:m-levels} mapped the double merge scenario to MARL approaches at various M levels.

To date, AD research has mostly focused on M0 with highly limited attempts at M1 and M2 such as those noted in Table \ref{table:m-levels}. We believe that a key reason is the lack of suitable AD simulation of interaction among heterogeneous traffic participants on the road.\footnote{Interestingly, multi-agent learning research at M3 through M5 happens more often for \textit{traffic management} \cite{cityflow,flow-berkeley,wei2019colight}, for which simulation has been under heavy development for about twenty years \cite{krajzewicz2002sumo, dresner2004aim,barcel2005micro}.} When it comes to MARL for AD interaction, the behavior of the simulated traffic participants must be both realistic and diverse, especially for the \textit{social vehicles} (i.e. vehicles sharing the driving environment with the autonomous vehicle) interacting with the \textit{ego vehicles} (i.e. vehicles controlled by the AD software).

To be sure, there are some excellent open-source simulators (see Appendix \ref{appendix:sims} for a survey), such as CARLA \cite{dosovitskiy2017carla} and SUMO \cite{krajzewicz2002sumo}, and even environments explicitly designed for RL and multi-agent research, such as Flow \cite{flow-berkeley} and AIM4 \cite{AIMAuton91:online}. Despite their respective successes, none of them comes close to addressing the specific need for in-depth MARL research on AD. The most specifically relevant efforts are probably highway-env \cite{highway-env}, which is architected as a set of independently hand-crafted interaction scenarios, and BARK \cite{bernhard2020bark}, which emphasizes the development of behavior models without explicit support for multi-agent research. Moreover, since there are only a small number of agent behavior models to choose from in the aforementioned solutions, it is still challenging to configure realistic interaction scenarios. Finally, the research community still lacks standard benchmarking suites that are comparable to MuJoCo \cite{todorov2012mujoco} in how it brought physics simulation to RL research and to StarCraft II \cite{vinyals2017starcraft} in how it brought large scale games to multi-agent research, but remain true to the AD reality. Consequently, papers were published with ad-hoc, one-off small scale simulations that are typically concerned with just a single task and typically not maintained or reusable. In response to such a situation, we propose \platform{} (\textbf{S}calable \textbf{M}ulti-\textbf{A}gent \textbf{R}L \textbf{T}raining \textbf{S}chool). Our goal is to provide an industrial-strength platform that helps to bring MARL research for AD to the next level and both empowers and challenges it for many years to come.

%% file: platform.tex
\section{Design Goals of the \platform{} Platform}

\begin{figure}[t!]
  \centering
  \includegraphics[width=0.95\textwidth]{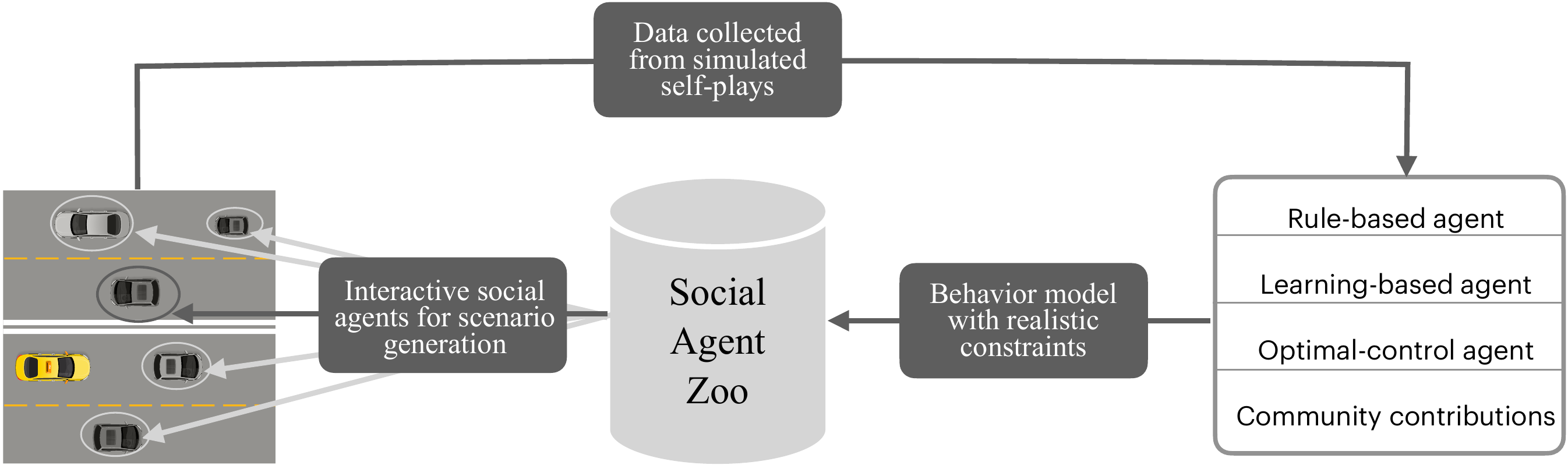}
  \caption{Bootstrapping interaction realism and diversity.}
  \label{figure:bootstrap}
\vspace{-10pt}
\end{figure}

%\subsection{Design Goals}
\paragraph*{Bootstrapping Realistic Interaction.} Realistic and diverse interaction arise out of the confluence of some key contextual factors: (1) \textit{physics}, (2) \textit{behavior of road users}, (3) \textit{road structure \& regulations}, and (4) \textit{background traffic flow}. The core of \platform{} covers all these aspects and is capable of continual growth to meet what realistic and diverse interaction requires. From the multi-agent perspective, the key to this continual growth is a bootstrapping strategy illustrated in Figure~\ref{figure:bootstrap}. At the heart is the \textit{Social Agent Zoo}, which collects social agents that are used to control social vehicles in \platform{} simulations. These social agents could come from self play \cite{silver2017mastering2} or population-based training \cite{vinyals2019alphastargrandmaster}. In order for the Social Agent Zoo to be grounded to the real world, however, at least some social agents will have to be also based on real-world driving data or domain knowledge. The process of growing the Social Agent Zoo is thus a multi-generational, iterative process that increases behavioral competence without sacrificing behavioral realism. As the Zoo grows, \platform{} will be able to simulate increasingly more complex but also increasingly more realistic multi-agent interaction. At the same time, given the open-source nature of \platform{}, we also expect that community contributions to the Social Agent Zoo will in turn benefit the global research community.

\paragraph*{Heterogeneous Agent Computing.}
Under the bootstrapping strategy, social agents could be as sophisticated as a full AD stack, or as simple as a few lines of scripts. They could use the same observations and actions as the ego agent running the AD stack, or take shortcuts to use simulator states directly. They could use more neural network modules than what the on-board vehicle computer could run, or they could use no neural networks at all. 
For SMARTS to rely on such social agents for realistic and diverse simulated behavior, it naturally has to accommodate agents that are heterogeneous in design and implementation. Moreover, for scalability, we may not want to always simulate social vehicle behavior at the highest fidelity. Instead, we may use compute-intensive agents only when and where fine-grained interaction matters, such as for unprotected left turns or double merges.
\begin{wraptable}{r}{8.1cm}
\footnotesize
\begin{tabular}{| m{1.5cm} | m{5.7cm} |}
    \hline
    \textbf{Area} & \textbf{Features} \\ \hline
    \multirow{6}{1.7cm}{Realistic Interaction} & Realistic physics \\
                                          & Heterogeneous ego and social agents \\
                                          & Handwritten social agents \\
                                          & Social agents trained with real-world data \\
                                          & Social agents trained with RL \\
                                          & Social agent zoo for crowd sourcing \\
    \hline
    \multirow{5}{1.7cm}{Platform Integration}  & Multi-agent distribution \\
                                          & Multi-simulation distribution \\
                                          & RLlib integration for RL training \\
                                          & SUMO integration for traffic simulation \\
                                          & Built-in scenario composition \\
    \hline
    \multirow{6}{1.7cm}{Research Friendliness} & Gym APIs \\
                                          & Headless mode \\
                                          & Web-based visualization with recording \\
                                          & Comprehensive observation \& action options \\
                                          & Multi-agent RL algorithm libraries \\
                                          & Realistic benchmark suits \\
    \hline
\end{tabular}
\caption{\platform{} features.}
\label{table:design-goals}
\vspace{-8mm}
\end{wraptable} 
In scenarios where fidelity of interaction does not matter as much, such as highway cruising without lane change, much simpler classical control agents may be used. 
To retain the flexibility to deploy diverse social agents where they best fit the need, \platform{} needs to be able to dynamically associate social vehicles with social agents that diverge in terms of vehicle model, observation and action spaces, and computing needs. To support this, we have devised a mechanism called ``bubbles'', which is discussed in Section \ref{section:scenarios} below.

\paragraph*{Key Features} To fully support research usage and to cover the major factors that underpin realistic and diverse interaction, \platform{} must also have additional features.We have identified the features in Table \ref{table:design-goals} as important and have provided support for them in the initial release of \platform{}. Here, we can only highlight a few key features from this list.
\begin{itemize}
    \item \platform{} is \textit{natively multi-agent}, by which we mean that social agents could be as intelligent as ego agents and as heterogeneous as need be.
    \item For scalable integration, in addition to distributed training of ego agents supported through Ray \cite{moritz2018ray}, \platform{} also manages its own \textit{distributed social agent computing}. % yaodong{should we briefly mentioned what communication protocol used for distributed computing, e.g. redis? ZMQ? etc. The idea is not only to say we support distirbuted, but also support it to the state-of-the-art.} JL: We currently go through sockets directly. This is actually pretty complicate stuff where we expect a major overhaul post open-sourcing.
    \item To provide the best possible \textit{researcher experience}, we follow the standard OpenAI Gym APIs, provide a web-streaming visualization solution, which allows ongoing simulation to be viewed from anywhere, such as on a smartphone, and offer a comprehensive set of observations and action spaces that can easily fit various research designs with minimum pre-processing and post-processing.
    \item Given our emphasis on support for MARL research, \platform{} integrates with popular libraries such as PyMARL \cite{samvelyan2019starcraft} and MAlib \cite{yingwenm12:online}, provides implementation of a few new algorithms, and supports a \textit{comprehensive set of MARL algorithms}. (A most comprehensive set to our knowledge. See Table~\ref{tab:algorithms} for specifics.)
    \item \platform{} offers a \textit{MARL benchmarking suite with AD-specific evaluation metrics}, which both poses challenging research questions and remains true to the AD reality. More benchmarking suites based on \platform{} are also in the workings.
    %and will be released gradually, with the first one coming out in 2020
\end{itemize}

\section{Architecture and Implementation}

To support our design goals, we have adopted a strongly \textit{compositional} approach that reflects the diverse sources of the complexity of multi-agent interaction in real-world driving. We use compositionality to both scale out for the size of the simulation and scale up for interaction complexity. Here, we can only comment on considerations about (1) how the simulation as a whole is composed, (2) how a specific interaction scenario is composed, and (3) how the various computational processes are orchestrated at runtime.

\begin{figure}[h]
  \centering
  \includegraphics[width=0.8\textwidth]{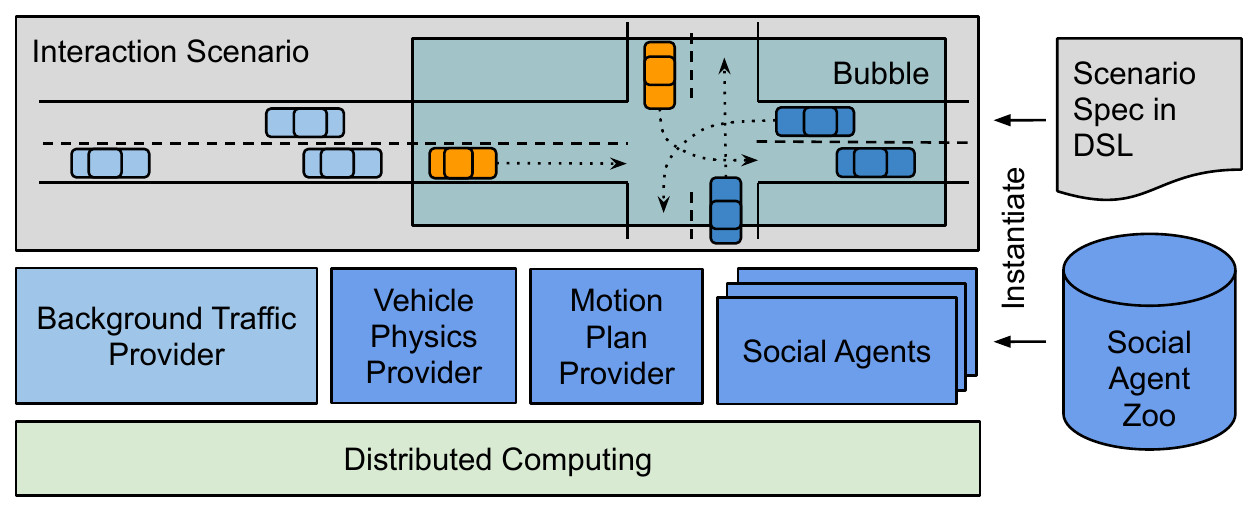}
  \caption{SMARTS Architecture. Interaction scenarios are defined with a domain-specific-language (DSL). Social agents are instantiated from the Social Agent Zoo. Orange vehicles are controlled by learning agents, dark blue vehicles by social agents, light blue ones by traffic provider. All providers and agents in principle run in their own processes, possibly remotely.}
  \label{figure:architecture}
  \vspace{-3mm}
\end{figure}

\paragraph*{Simulation Providers.}
We adopt a \textit{provider architecture}, which use limited co-simulators, or providers, for the major factors underpinning interaction. The overall simulation step orchestrates the stepping of these providers.
\textit{Background traffic flow} is provided everywhere in the map during the simulation. The map could be as big as a city and as small as an intersection. The background traffic provider is responsible for bringing the simulation session to the point where ``foreground'' interaction of interest could happen, at which point, the evolution of the traffic will be determined through the interaction of the social agents. \platform{} currently supports SUMO \cite{krajzewicz2002sumo} as a background traffic provider.\footnote{CARLA recently added trial integration with SUMO. This is a welcome move. Through our traffic provider interface, \platform{} may also be adapted to work with Aimsun Next \cite{casas2010traffic} or Berkeley Flow \cite{flow-berkeley}.}
\textit{Physics} is supported through the \textit{vehicle provider}, which simulates the specified physical vehicle model and exposes a control interface at different levels of abstraction: throttle-brake-steering, trajectory tracking, and lane following. With easily replaceable vehicle models and tire models, investigation on how physical changes, such as wet road surface or a flat tire, could affect multi-agent interaction may be readily studied. Our current implementation utilizes the \emph{Bullet physics engine} \cite{BulletRe59} and supplements it with vehicle-specific dynamics and controllers. Finally, the \textit{motion plan provider} can be used to implement social agents dedicated to highly specific maneuvers such as cut-in or U-turn. Our implementation includes three options: motion models available through SUMO, a simple Bezier-curve based motion planner \cite{scheiderer2019bezier}, and an interactive motion planner based on OpEn \cite{open2020}.

\paragraph*{Interaction Scenarios.}\label{section:scenarios}
\begin{wrapfigure}{r}{0.5\textwidth}
  \hspace{2mm}
  \vspace{-3mm}
  \begin{center}
  \includegraphics[width=0.45\textwidth]{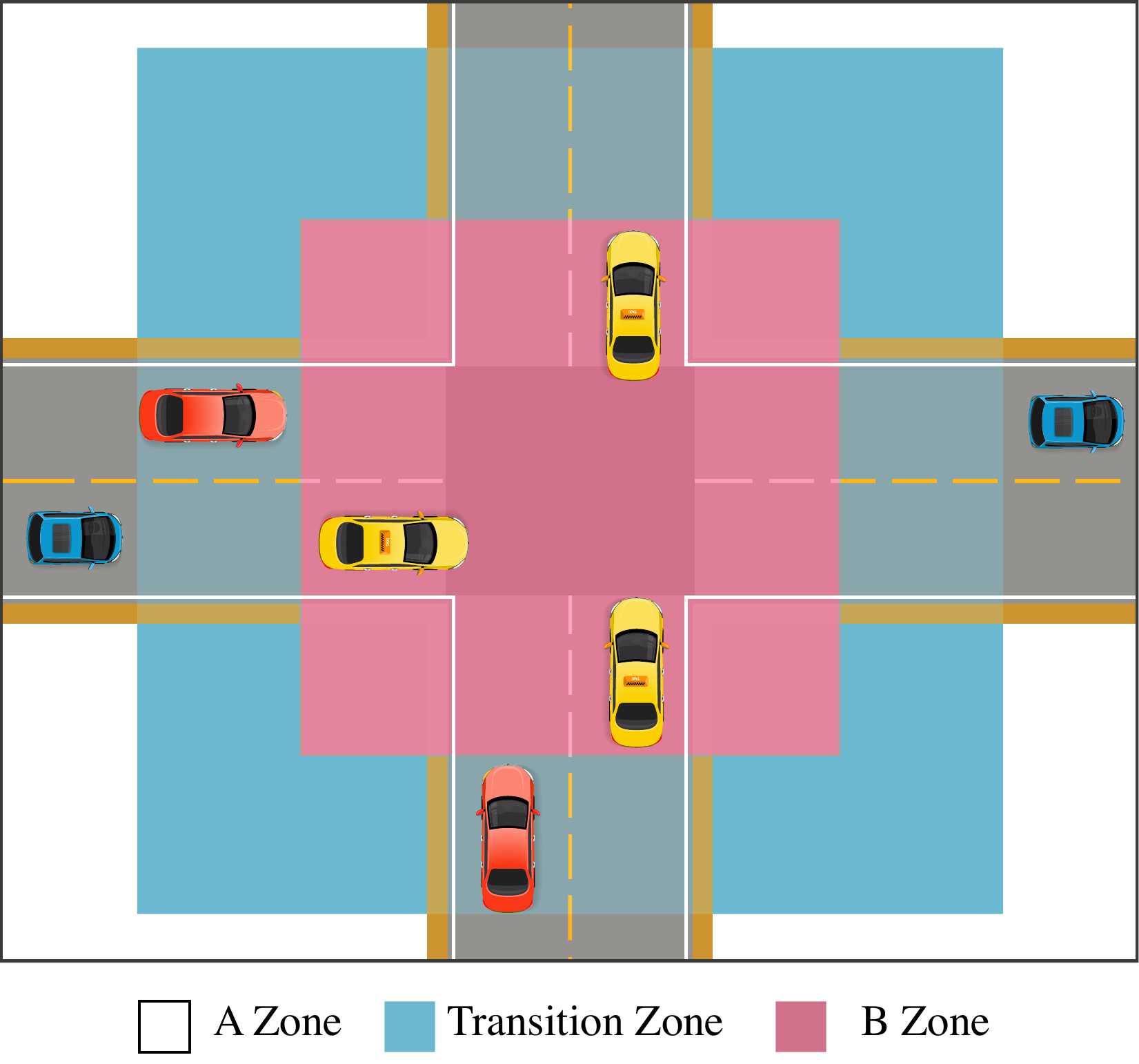}
  \caption{Bubble. Inside A Zone, social vehicles are controlled by the traffic provider; inside B Zone, by agents from the Social Agent Zoo.}
  \label{figure:bubble}
  \end{center}
  \vspace{-1mm}
\end{wrapfigure}
To support large scale compositional construction of interaction scenarios, \platform{} comes with a Python-based domain specific language (DSL). At the base level, a scenario is composed of a map, a set of routes through the map, a set of vehicles with various driving characteristics, and ``flows'', which are assignments of the vehicles to the routes. As this level, we essentially just use the DSL to preprogram the background traffic. To integrate agents in the Social Agent Zoo into a scenario, we use ``bubbles''. 
A bubble is a spatiotemporal and conditionally specifiable region in which social vehicles are expected to be controlled by agents from the zoo. Different from \textit{ObservedWorld} \cite{bernhard2020bark}, which is used for planning or opponent modeling, bubble is for interaction management. As the background traffic (managed by the traffic provider) hits the outer membrane of the bubble, a transition happens wherein the control of the social vehicles could be handed over to specific social agents.
To facilitate this handover, the bubble logic will need to instantiate the necessary vehicle and sensor models,\footnote{This is necessary because the background traffic provider typically models the social vehicles as mere rectangles with no articulated wheels or steering mechanism.} links these to the social agent's observation and action interfaces, starts the social agent processes, and switch control over from the traffic provider to the social agents.
Imagine a large road network with many unprotected left turns and double merges, bubbles can be placed at these places wherein the multi-agent interaction can be configured and studied in detail. Accordingly, ego agents could be trained in a highly focused way by only collecting interaction trajectories in these bubbles, while the traffic provider continues to supply realistic traffic flow through the outer membrane of the bubble. This is how the bubble orchestrates the providers, the Social Agent Zoo, and the computing resource to create a milieu that is rich in multi-agent interaction.

\paragraph*{Distributed Computing}\label{section:computation}
The bubble mechanism allows \platform{} to scale up without sacrificing interaction realism. Since the social agents may require a lot of compute, we cannot treat them as ``non-character-players'' (NPCs) run by simple scripts. Instead, we need to elastically assign computing resource according to the need of simulation fidelity. Another issue here is managing the dependencies of the social agents, especially their deep learning dependencies. Through the bootstrapping process, these agents are likely going to be based on more and more sophisticated deep models. \platform{} generalizes the approach Ray \cite{moritz2018ray} takes to run scalable distributed training of ego agents by making social agents run in their own processes, possibly on remote machines, and provide options for running them with or without Ray and with or without TensorFlow and PyTorch dependencies. This way, we can run large scale, possibly city-scale, multi-agent simulations wherein one computer is dedicated to running the background traffic of the entire city, while many other computers are dedicated to running social agents inside the various bubbles.\footnote{To get an intuitive sense of what SMARTS simulations are like when all these are up and running,
please see sample screenshots in Appendix~\ref{appendix:smarts-demos}.}

%% file: benchmarking.tex
\section{Support for  Multi-Agent Reinforcement Learning}\label{section:benchmarking}

% TODO by ZM A: given a set of MARL algorithms, what kind of experimental results we can produce? How can we compare the different algorithms? 

\begin{figure}[ht]
    	\centering
    	\begin{subfigure}[b]{0.24\textwidth}
    		\centering
    		\includegraphics[width=\textwidth]{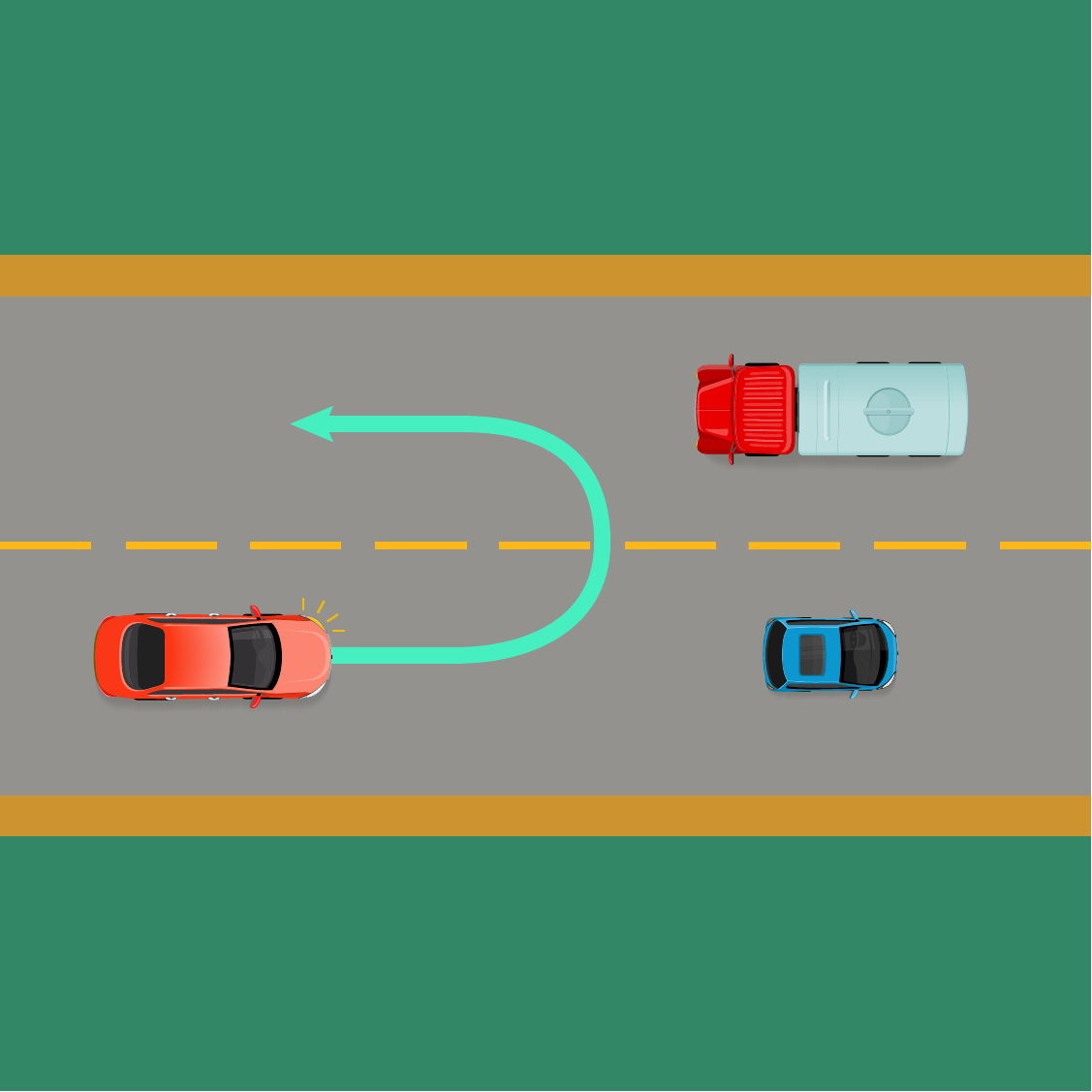}
    		\subcaption{U-turn}
    		\label{subfig:u-turn}
    	\end{subfigure}
    	\begin{subfigure}[b]{.24\textwidth}
    		\centering
    		\includegraphics[width=\textwidth]{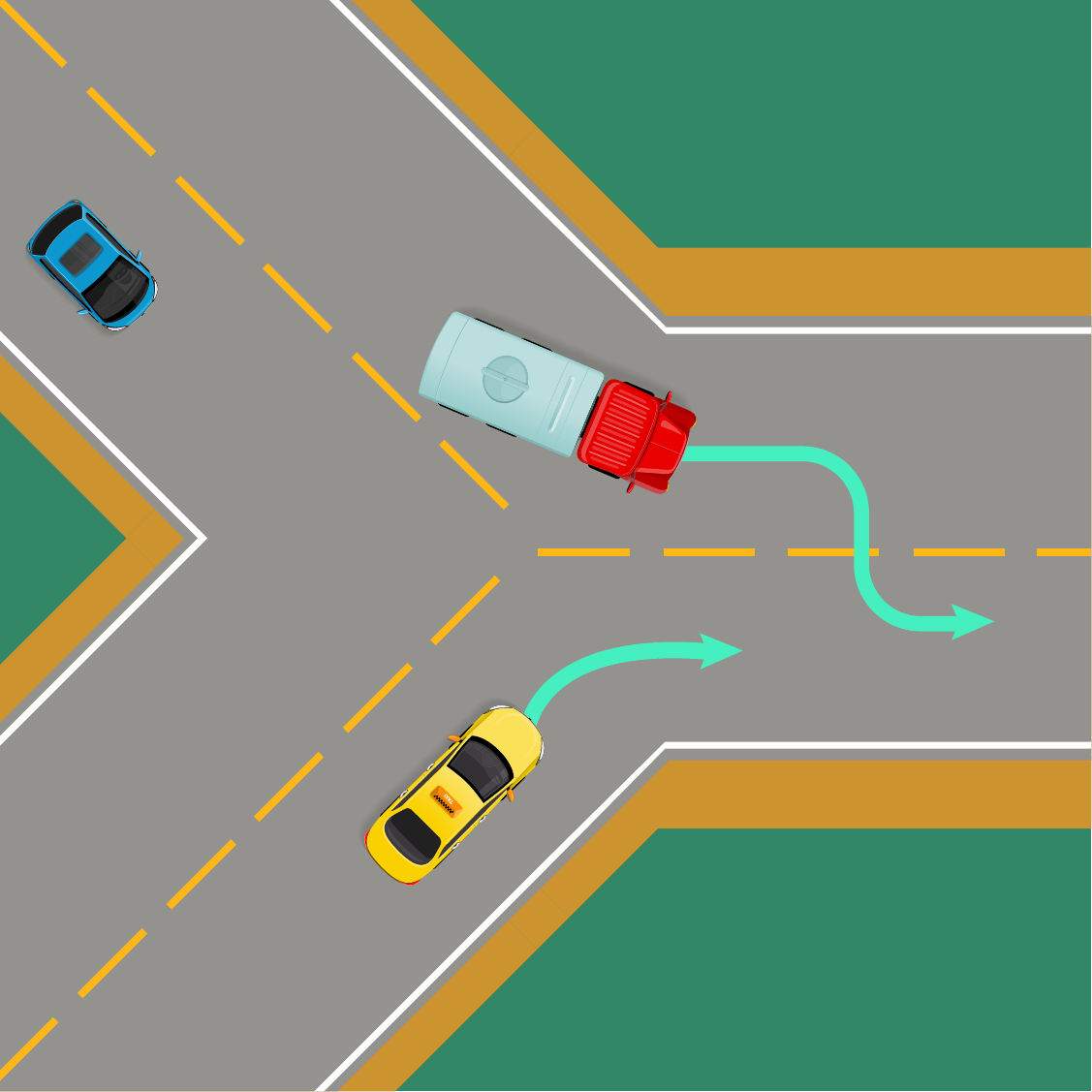}
    		\subcaption{Lane merging}
    		\label{subfig:lane-merging}
    	\end{subfigure}
    	\begin{subfigure}[b]{0.24\textwidth}
    		\centering
    		\includegraphics[width=\textwidth]{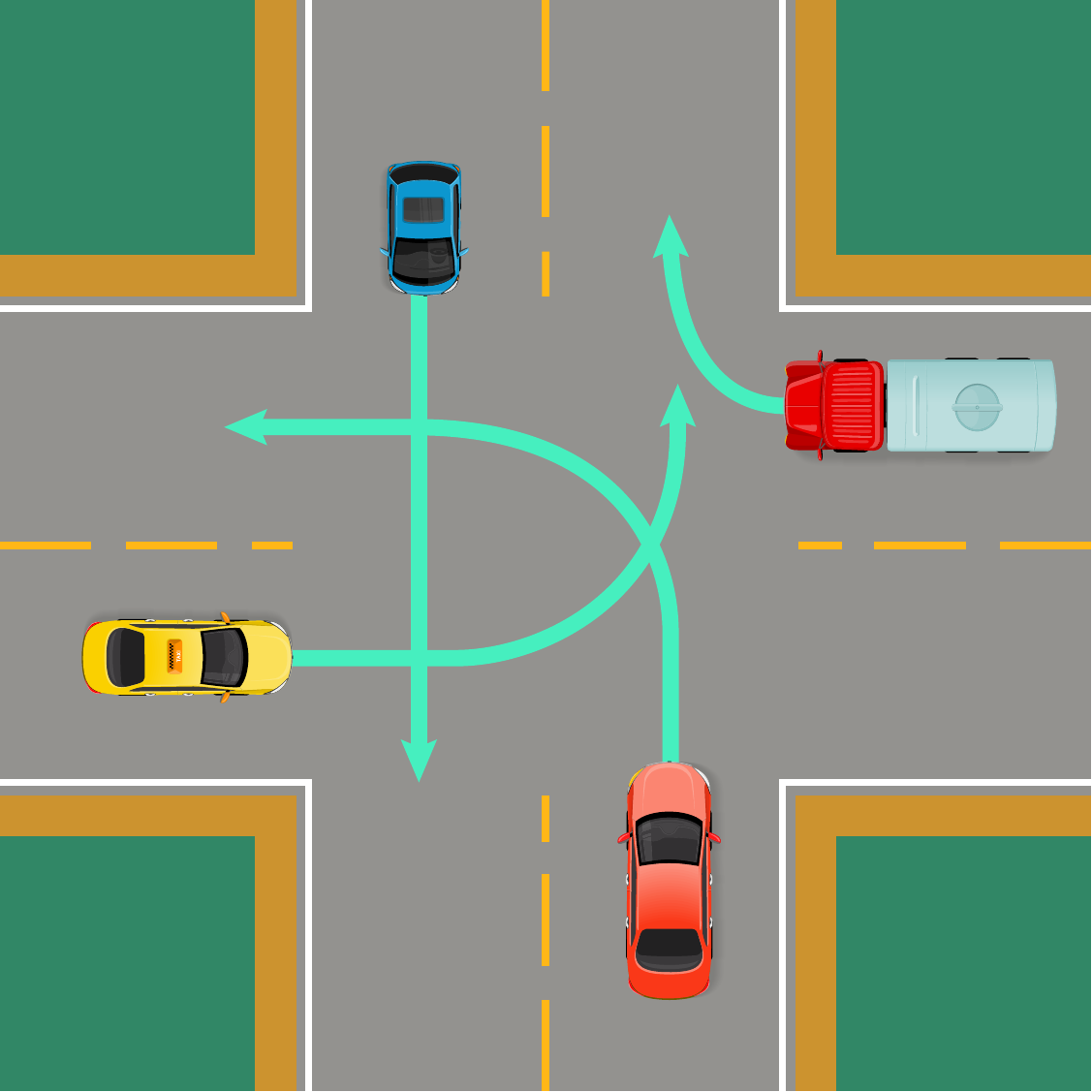}
    		\subcaption{Intersection}
    		\label{subfig:intersection}
    	\end{subfigure}
    	\begin{subfigure}[b]{0.24\textwidth}
    		\centering
    		\includegraphics[width=\textwidth]{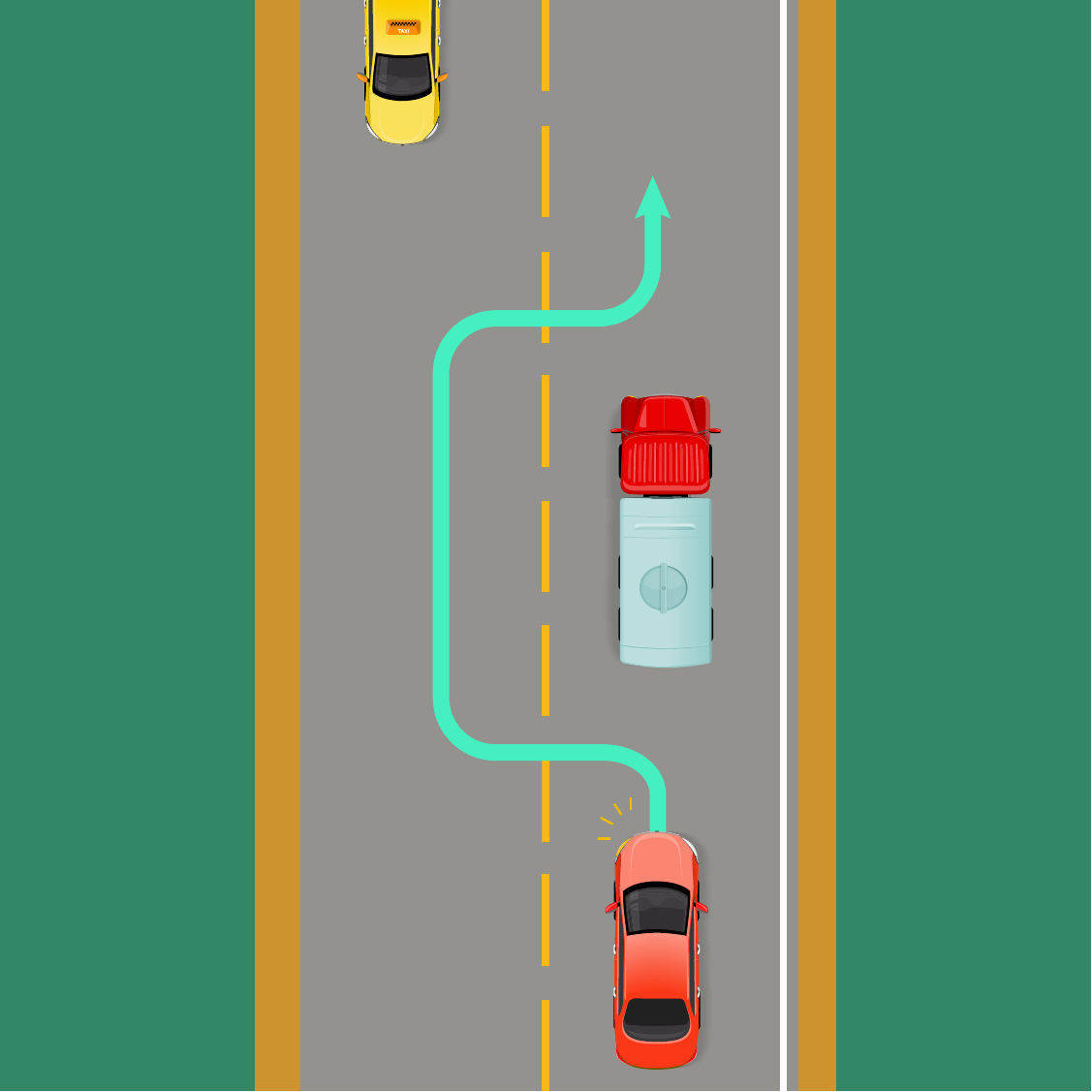}
    		\subcaption{Overtaking}
    		\label{subfig:overtaking}
    	\end{subfigure}
    	\begin{subfigure}[b]{0.24\textwidth}
    		\centering
    		\includegraphics[width=\textwidth]{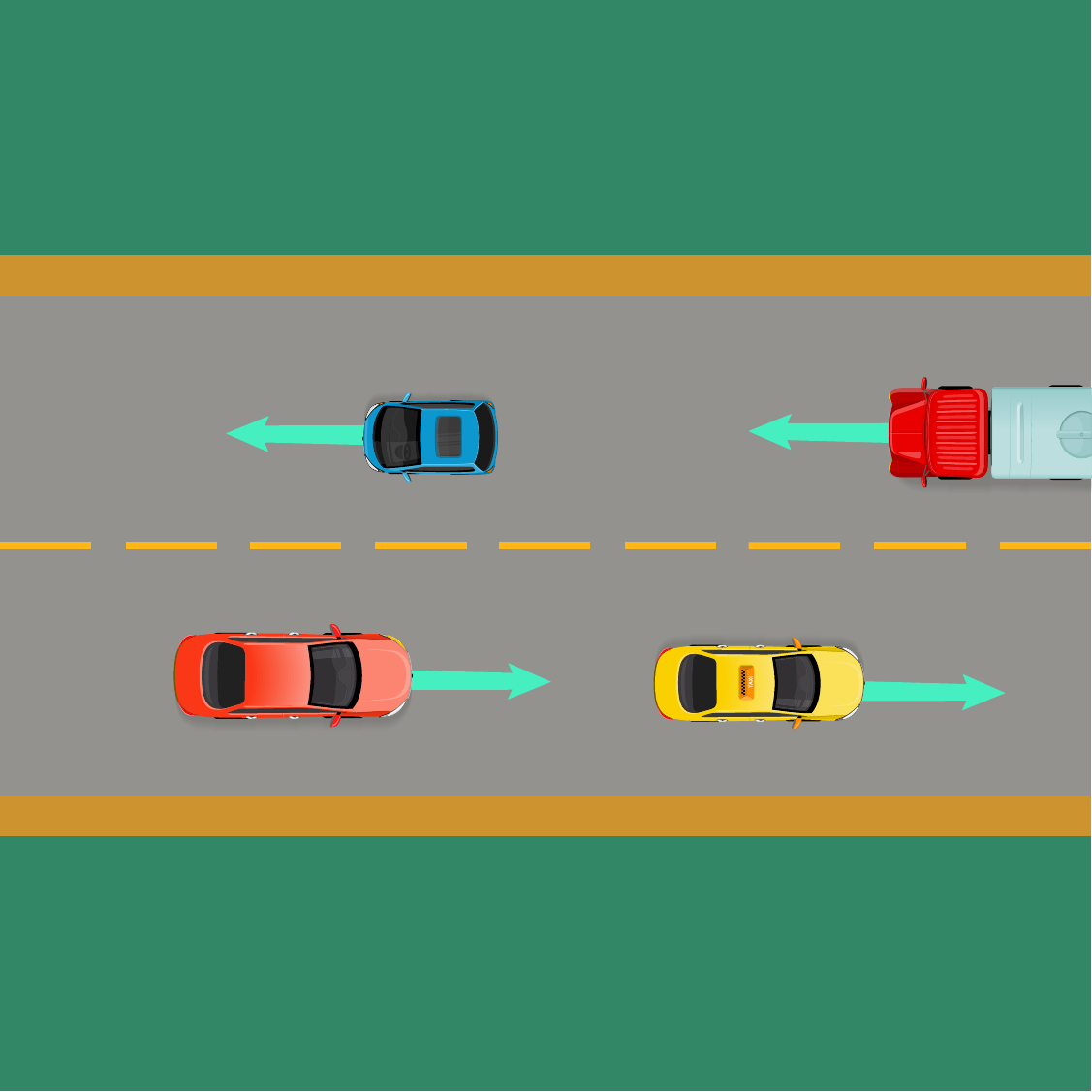}
    		\subcaption{Two-way traffic}
    		\label{subfig:two-way}
    	\end{subfigure}
    	\begin{subfigure}[b]{0.24\textwidth}
    		\centering
    		\includegraphics[width=\textwidth]{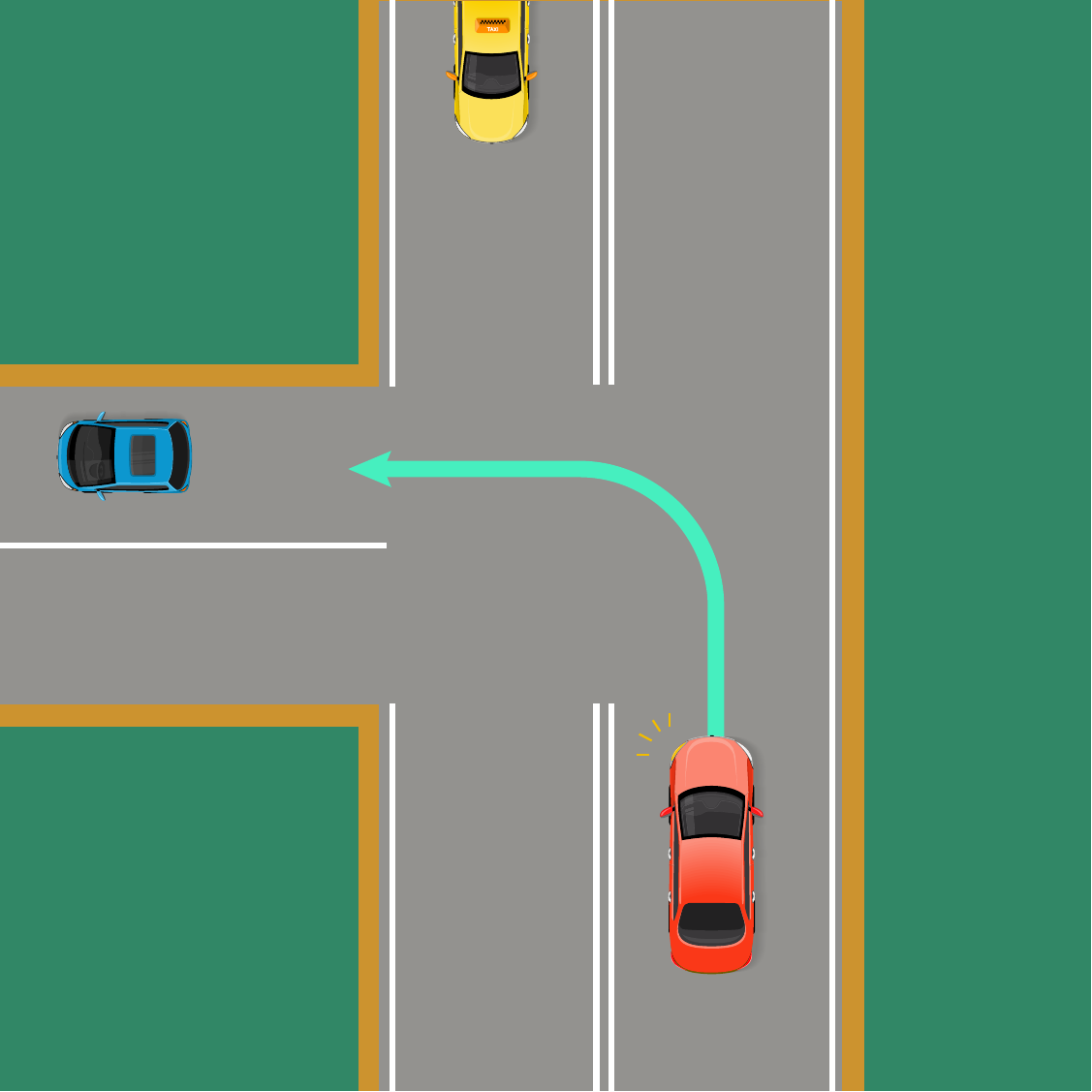}
    		\subcaption{Unprotected left turn}
    		\label{subfig:left-turn}
    	\end{subfigure}
    	\begin{subfigure}[b]{0.24\textwidth}
    		\centering
    		\includegraphics[width=\textwidth]{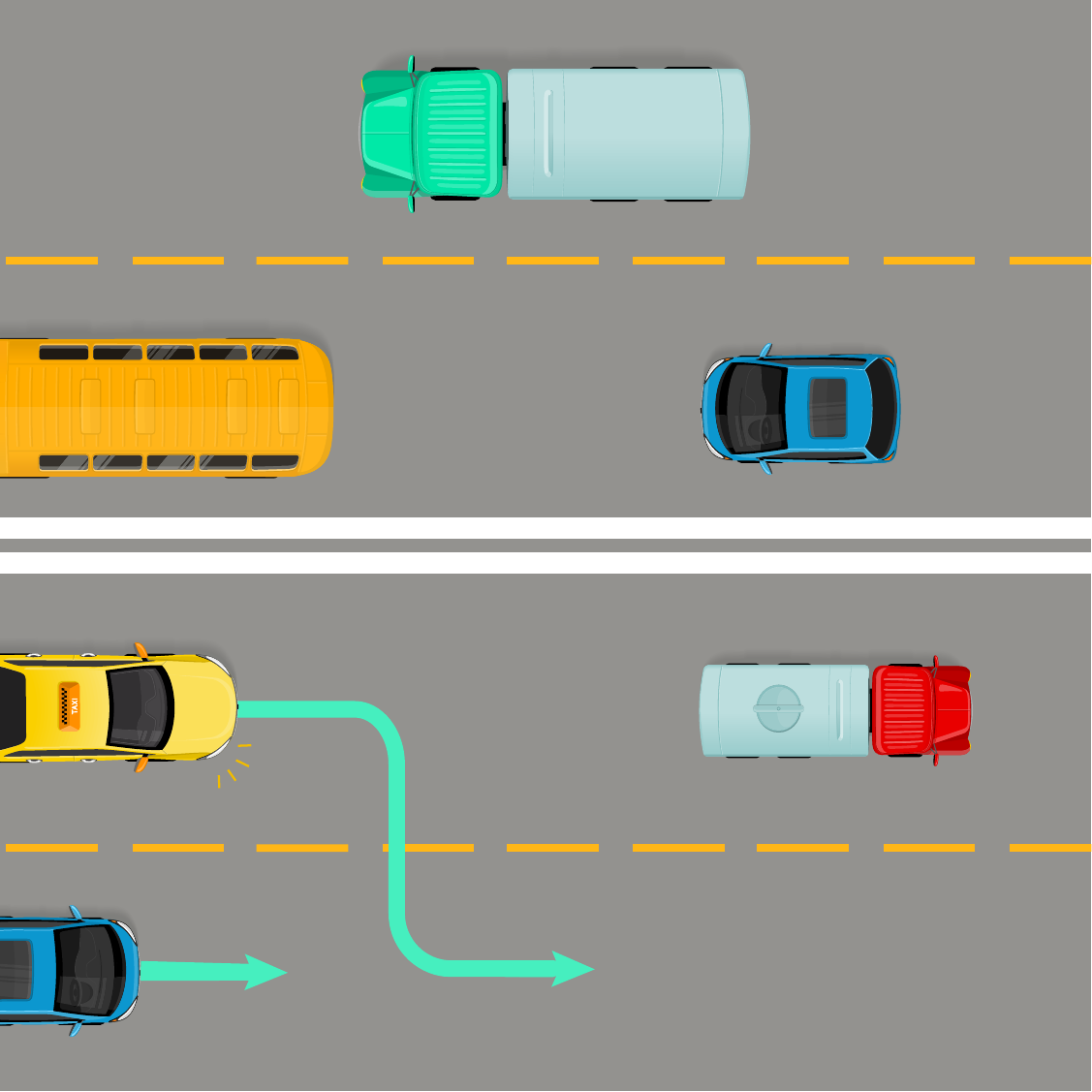}
    		\subcaption{Cut-in}
    		\label{subfig:cut-in}
    	\end{subfigure}
    	\begin{subfigure}[b]{0.24\textwidth}
    		\centering
    		\includegraphics[width=\textwidth]{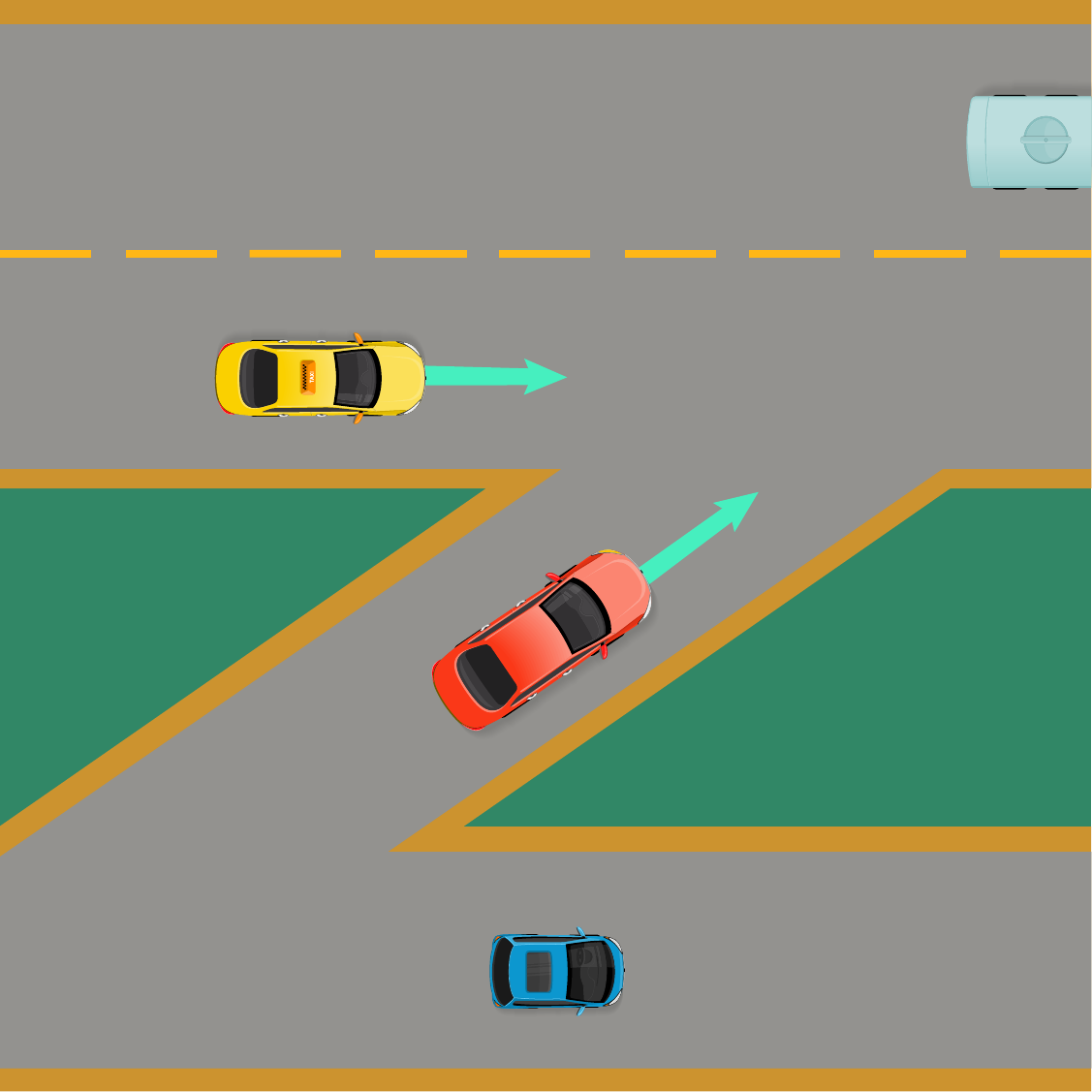}
    		\subcaption{On-ramp merge}
    		\label{subfig:on-ramp}
    	\end{subfigure}
    	\begin{subfigure}[b]{0.24\textwidth}
    		\centering
    		\includegraphics[width=\textwidth]{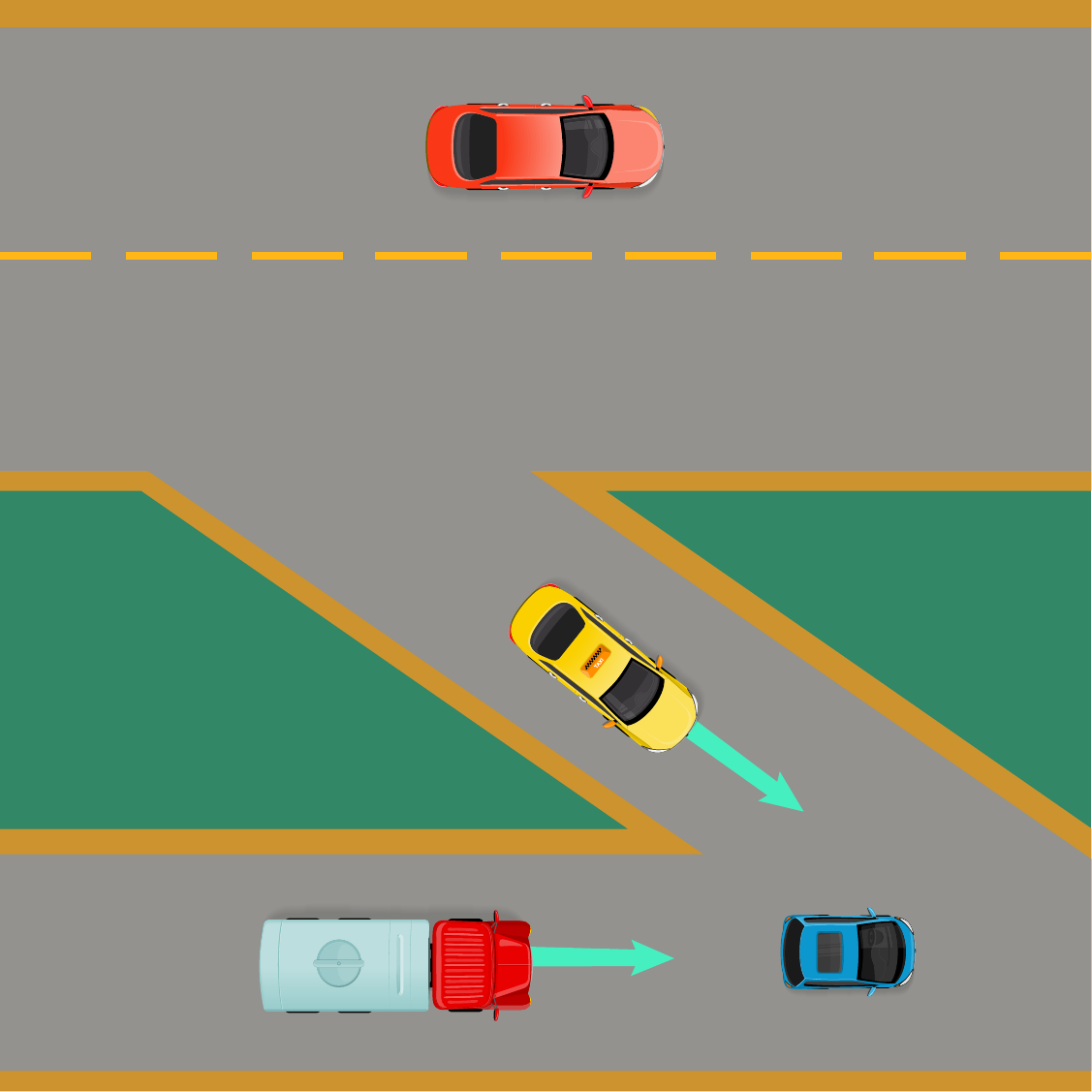}
    		\subcaption{Off-ramp merge}
    		\label{subfig:off-ramp}
    	\end{subfigure}
    	\begin{subfigure}[b]{0.24\textwidth}
    		\centering
    		\includegraphics[width=\textwidth]{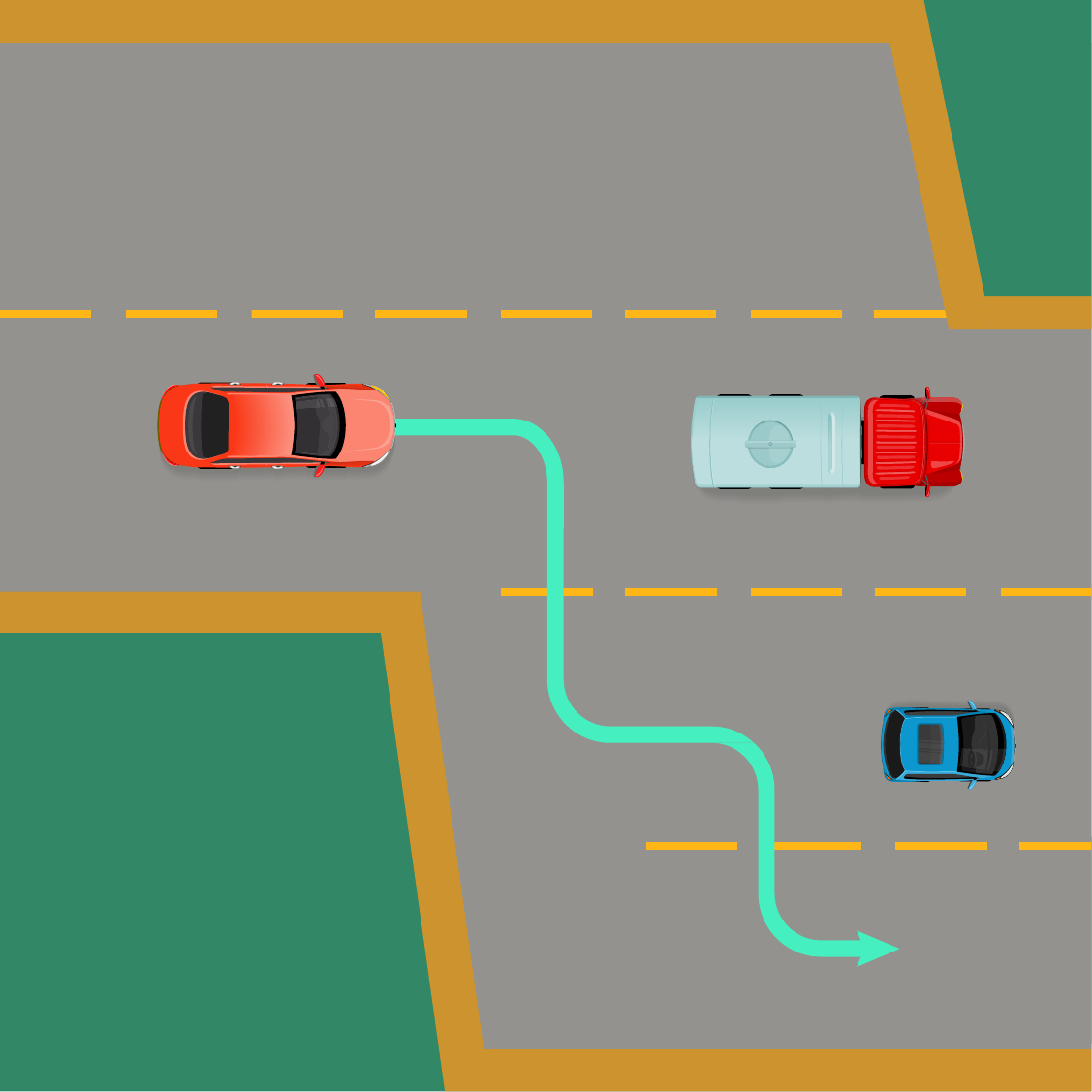}
    		\subcaption{Cascading lane change}
    		\label{subfig:cascading-lane-change}
    	\end{subfigure}
    	    	\begin{subfigure}[b]{0.24\textwidth}
    		\centering
    		\includegraphics[width=\textwidth]{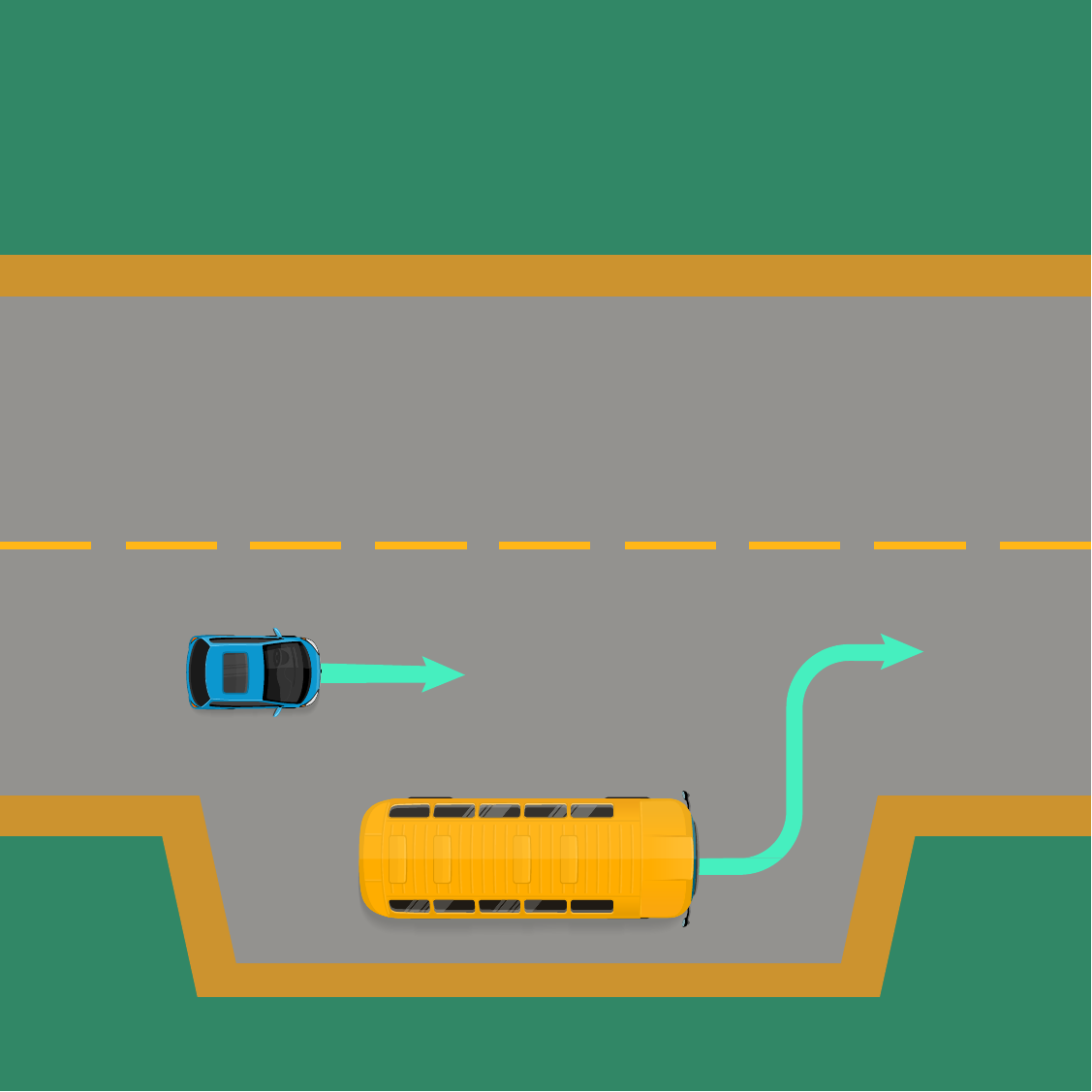}
    		\subcaption{Bus stop merge}
    		\label{subfig:bus-stop}
    	\end{subfigure}
    	    	\begin{subfigure}[b]{0.24\textwidth}
    		\centering
    		\includegraphics[width=\textwidth]{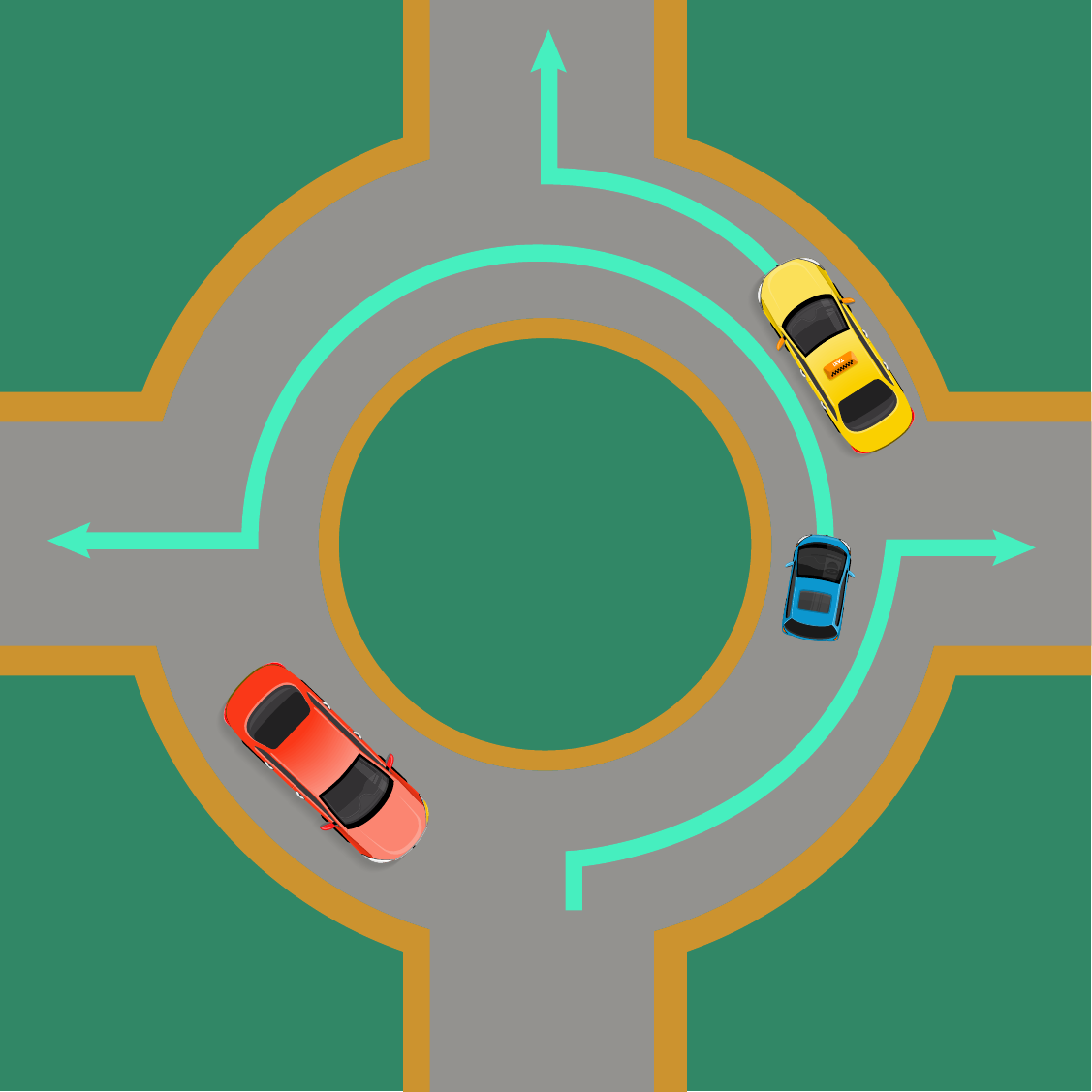}
    		\subcaption{Roundabout}
    		\label{subfig:roundabout}
    	\end{subfigure}
    	\caption{Scenarios of driving interaction specifiable in \platform{}}
    	\label{fig:all-scenarios}
    \end{figure}

    Using the scenario DSL of \platform{}, we can create numerous scenarios that vary in road structure and traffic. Figure~\ref{fig:all-scenarios} highlights some of them. These scenarios provide rich traffic flows and road conditions to help us study behavior and driving strategies. To train and evaluate the ego agents with specific parameters in these scenarios, we have implemented a benchmark runner based on Ray \cite{moritz2018ray} and RLlib \cite{liang2018rllib}. With the support of Ray, \platform{} enables users to conduct scalable and parallelizable training and evaluations. Below, we briefly describe other key elements of \platform{} for supporting MARL research.

    \paragraph*{Observation, Action and Reward.} In \platform{}, the observation space for an agent is specified as a configurable subset of available sensor types that include dynamic object list, bird's-eye view occupancy grid maps and RGB images, ego vehicle states, and road structure etc. To make the use of deep models easier, we also provide built-in adapters to convert sensor data to tensors. An agent's action space is determined by the controller chosen for it. \platform{} supports four types of controllers: \texttt{ContinuousController}, \texttt{ActuatorDynamicController}, \texttt{TrajectoryTrackingController} and \texttt{LaneFollowingController}. The mapping from controllers to action spaces is listed in the Appendix~\ref{appendix:controllers}. For reward, \platform{} provides distance travelled along the mission route per time step as the raw reward signal, which may be used in reward customization that combines feedback on key events (e.g. collision, off-road, etc.) and other observations. % \ming{maybe we can show more details} % \weinan{move the code to appendix.} % Below is a piece of code to show the subscription of observation features.
    
    % \begin{verbatim}
    %     # subscribe observation features, support multiple 
    %     # sensor data subscription ...
    %     subscription = dict(goal_relative_pos=True, 
    %         heading_errors=True, neighbor=True, img_gray=True)
    %     observation_space = gym.spaces.Dict(
    %         common.subscribe_features(**subscription)
    %     )
    % \end{verbatim}

    \paragraph*{Algorithms \& Baselines.} \platform{} integrates three popular (MA)RL libraries---RLlib \cite{liang2018rllib}, PyMARL \cite{samvelyan2019starcraft}, and MAlib \cite{yingwenm12:online}, allowing many algorithms to be used directly for AD research. While our experiments reported below used RLlib for high-concurrency training, RLlib's support for MARL is highly limited, because it focuses mainly on the implementation of single agent reinforcement learning. We therefore have provided implementations of more MARL algorithms under additional paradigms, such as centralized training \& decentralized execution (CTDE) and networked agent learning \cite{zhang2018fully}. Appendix~\ref{appendix:algorithms} overviews the algorithms currently available in \platform{}. To our knowledge, it is by far the most comprehensive set of MARL algorithms, covering all MARL paradigms (as show in Figure~\ref{figure:marl-paradigm}).
    %\yaodong{@ming, conclude this section with some summary statistics, say this is by far the most comprehensive lirary that support XXX MARL algorithms, and we are keep developing it.}
    
    \paragraph*{Metrics.} Existing evaluation metrics in autonomous driving tend to be limited to specific tasks \cite{bernhard2020bark,dosovitskiy2017carla}. Specifically, they focus on macro driving performance, such as \textit{1) Whether the agent reached its goal; 2) Collision ratio; 3) Distance from the agent to its goal; 4) Ratio of driving the wrong way}. They hardly explain why an agent achieves a certain performance level. Under the multi-agent setting, an agent's performance is not only a matter of how it itself behaves, but also depends on how other agents interact with it. Thus, for multi-agent evaluation in \platform{}, we propose a more comprehensive set of metrics in Table~\ref{tab:metrics}, covering three main aspects that include model performance, agent population behavior distribution, and game theoretic analysis.
    
    \begin{table}[h]
        \begin{center}
        \resizebox{0.9\columnwidth}{!}{
            \begin{tabular}{| c | c | c |}
                \hline
                \textbf{Metric} & \textbf{Type} & \textbf{Description} \\ \hline
                Collision Rate & Performance & \# of collisions over \# of episodes. \\
                \hline
                Completion Rate & Performance & \# of missions completed over \# of episodes. \\
                \hline
                Generalization & Performance & Robustness of algorithms to scenario variation. \\
                \hline
                Safety & Behavior & Integrated metrics, e.g. non-collision rate. \\
                \hline
                Agility & Behavior & Integrated metrics, e.g. speed. \\
                \hline
                Stability & Behavior & Integrated metrics for driving smoothness. \\
                \hline
                Control Diversity & Behavior & Preference for longitudinal or lateral control. \\
                \hline
                Cut-in Ratio & Behavior & Probability of cut-in in traffic flow. \\
                \hline
                Stochasticity & Behavior & Stochasticity of decision making. \\
                \hline
                Collaborative & Game theory & Compatible interests, e.g. ratio of giving way. \\
                \hline
                Competitive & Game theory & Conflicting interests, e.g. ratio of overtaking. \\
                \hline
                % Competitive and Collaborative Mixed & Game theory & \\
                % \hline
            \end{tabular}}
        \end{center}
        \caption{Evaluation metrics for multi-agent autonomous driving. User can find the implementation of them in the \texttt{Metrics} class located in benchmarking library of \platform{}. Also, it is very easy for users to extend these metrics by inheriting from the \texttt{Metrics} class.}
        \label{tab:metrics}
        \vspace{-1mm}
    \end{table}

%% file: experiments.tex
\section{Experiments \& Results}
To demonstrate how \platform{} may support MARL research on AD, we run a series of experiments that involve using seven MARL algorithms to tackle three increasingly more challenging driving scenarios that require non-trivial interactive capabilities to be learned. 
    \begin{itemize}
        \item Two-Way traffic: As Figure~\ref{subfig:two-way} shows, agents start from either of two opposite lanes, then drive straight to the other end without colliding into each other or into social vehicles.
        \item Double Merge: As illustrated in Figure~\ref{fig:double-merge} and discussed in the Introduction.
        \item Unprotected Intersection: As Figure~\ref{subfig:intersection} shows, agents will start from any of the four roads linked to a junction, pass through the junction, and then continue on to other roads.
    \end{itemize}
Each agent in each scenario has a different mission to complete: driving from a specific start to a specific goal. Traffic consists of the agents in training as well as the background traffic provided by SUMO \cite{krajzewicz2002sumo}. The chosen baselines include two independent learning algorithms, DQN \cite{mnih2013playing} and PPO \cite{schulman2017proximal}, four centralized training methods, MAAC, MF-AC \cite{yang2018mean}, MADDPG\cite{lowe2017multi}, and Networked Fitted-Q \cite{zhang2018networked}, and a fully centralized method, CommNet \cite{sukhbaatar2016learning}. The observation, action, and reward functions are kept the same for all these baselines.

\paragraph{Observation.} The observation is a stack of three consecutive frames, which covers the dynamic objects and key events. For each frame, it contains: \textit{1) relative position of goal; 2) distance to the center of lane; 3) speed; 4) steering; 5) a list of heading errors; 6) at most eight neighboring vehicles' driving states (relative distance, speed and position); 7) a bird's-eye view gray-scale image with the agent at the center.}
\paragraph{Action.} The action used here is a four-dimensional vector of discrete values, for longitudinal control---\textit{keep lane} and \textit{slow down}---and lateral control---\textit{turn right} and \textit{turn left}.
\paragraph{Reward.} The reward is a weighted sum of the reward components shaped according to ego vehicle states, interactions involving surrounding vehicles, and key traffic events. More details can be found in our implementation code.

\paragraph*{Performance.} Table~\ref{tab:cc-all} shows the results of the baselines. It reports average \textit{collision rate} and \textit{completion rate} of the agent population on 10 episodes under two different background traffic settings. Unsurprisingly, all algorithms score higher under the No Social Vehicle setting. MADDPG outperforms the others in most tasks, especially in Intersection. Since the reward function and the observation are shared among all algorithms, independent learning agents could be at a disadvantage. For example, the better performance of MADDPG in the toughest setting of unprotected intersection, we believe, could be explained by the fact that agents learn to interact better by leveraging extra information from other agents. Since the overall interaction dynamics of the environment depends on all the agents, and the driving behavior of one agent will be influenced by other agents, extra information from other agents will mitigate the uncertainty for decision making, and lead the policy learning process into a more reasonable direction. (See Table~\ref{tab:cc-big} for additional results.) %\yaodong{@ming, should we put table 7 in the main text? I don;t see why a comprehensive list of performance shuld stay in appendix}

    \begin{table}[h]
        \renewcommand\arraystretch{1.2}
        \begin{center}
        \resizebox{0.85\columnwidth}{!}{
            \begin{tabular}{ccccccc}
            \toprule
            \multirow{2}{*}{Scenario} & \multicolumn{3}{c}{No Social Vehicle} & \multicolumn{3}{c}{Random Social Vehicle} \\
                     & PPO & CommNet & MADDPG & PPO & CommNet & MADDPG \\ \midrule
            Two-Way & \textbf{0/1} & 0/0.96 & \textbf{0/1} & 0.25/0.75 & 0.25/0.65 & \textbf{0.13/0.87} \\ 
            % Double Merge & 0.8/0.33 & 0.05/0.05 & \textbf{0.05/0.95}  & 0.85/0.125   & 0.375/0.45  & &
            % \textbf{0.4/0.6}
            Double Merge & \textbf{0/1} & 0.7/0.25 & 0.1/0.9 & \textbf{0.02/0.98} & 0.5/0.5 & 0.17/0.8 \\
            Intersection & 0.1/0.07 & 0.3/0.7 & \textbf{0/1} & 0.50/0.45 & 0.5/0.45 & \textbf{0.30/0.7}  \\ \bottomrule
            \end{tabular}
        }
        \end{center}
        \caption{Average Collision Rate / Completion Rate of three baselines on goal-directed navigation tasks in three scenarios. For collision rate, lower is better; for completion rate, higher is better. Collision and completion rates do not always sum up to 1, because agents may fail to reach destination before an episode ends. Examination of replay confirmed this for CommNet in Two-Way.}
        \label{tab:cc-all}
    \end{table}

\begin{figure}[ht]
    	\centering
    	\begin{subfigure}[b]{0.32\textwidth}
    		\centering
    		\includegraphics[width=\textwidth]{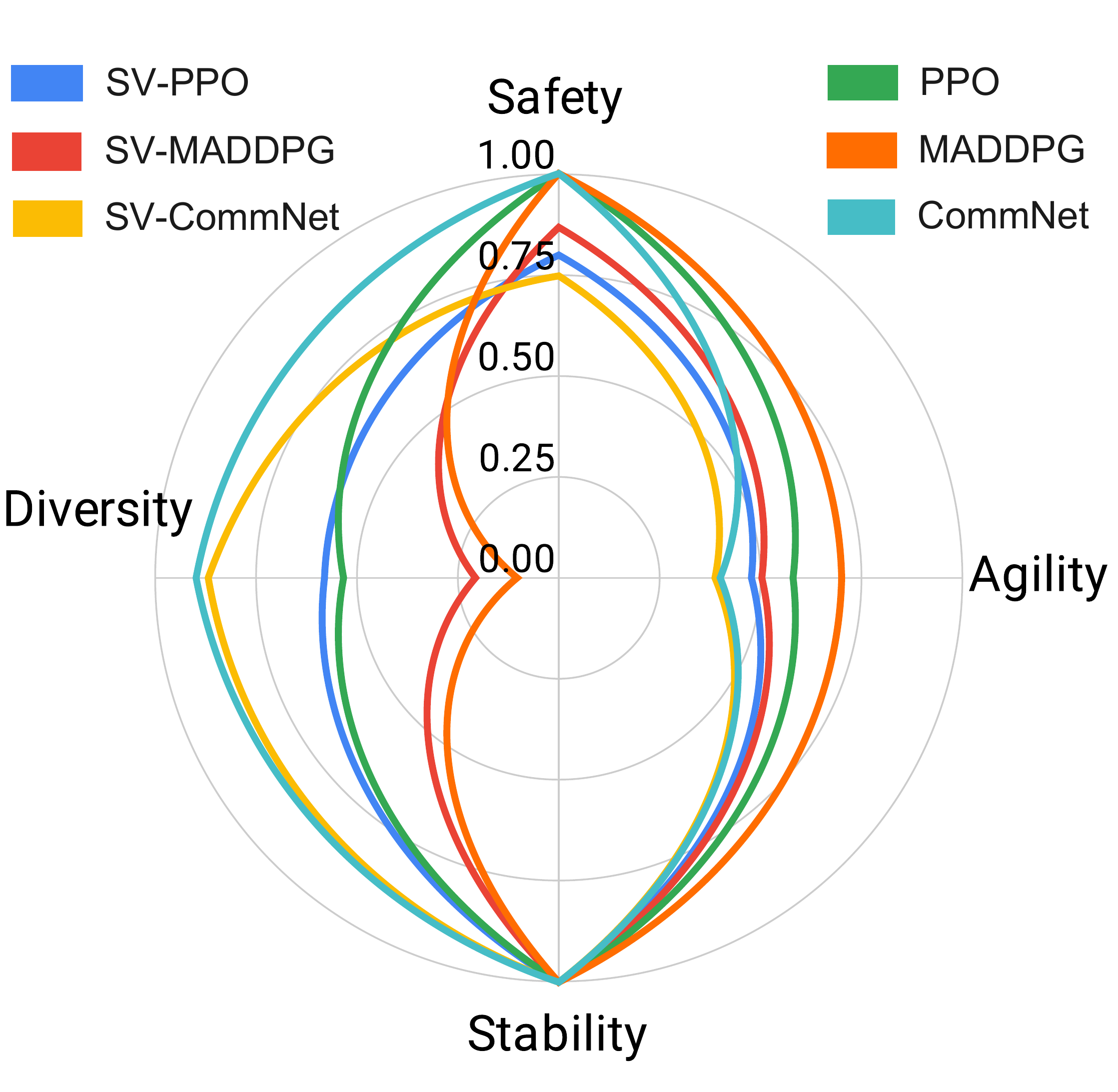}
    		\subcaption{Two-Way}
    	\end{subfigure}
    	\begin{subfigure}[b]{.317\textwidth}
    	    \setlength{\abovecaptionskip}{0.3cm} 
    		\centering
    		\includegraphics[width=\textwidth]{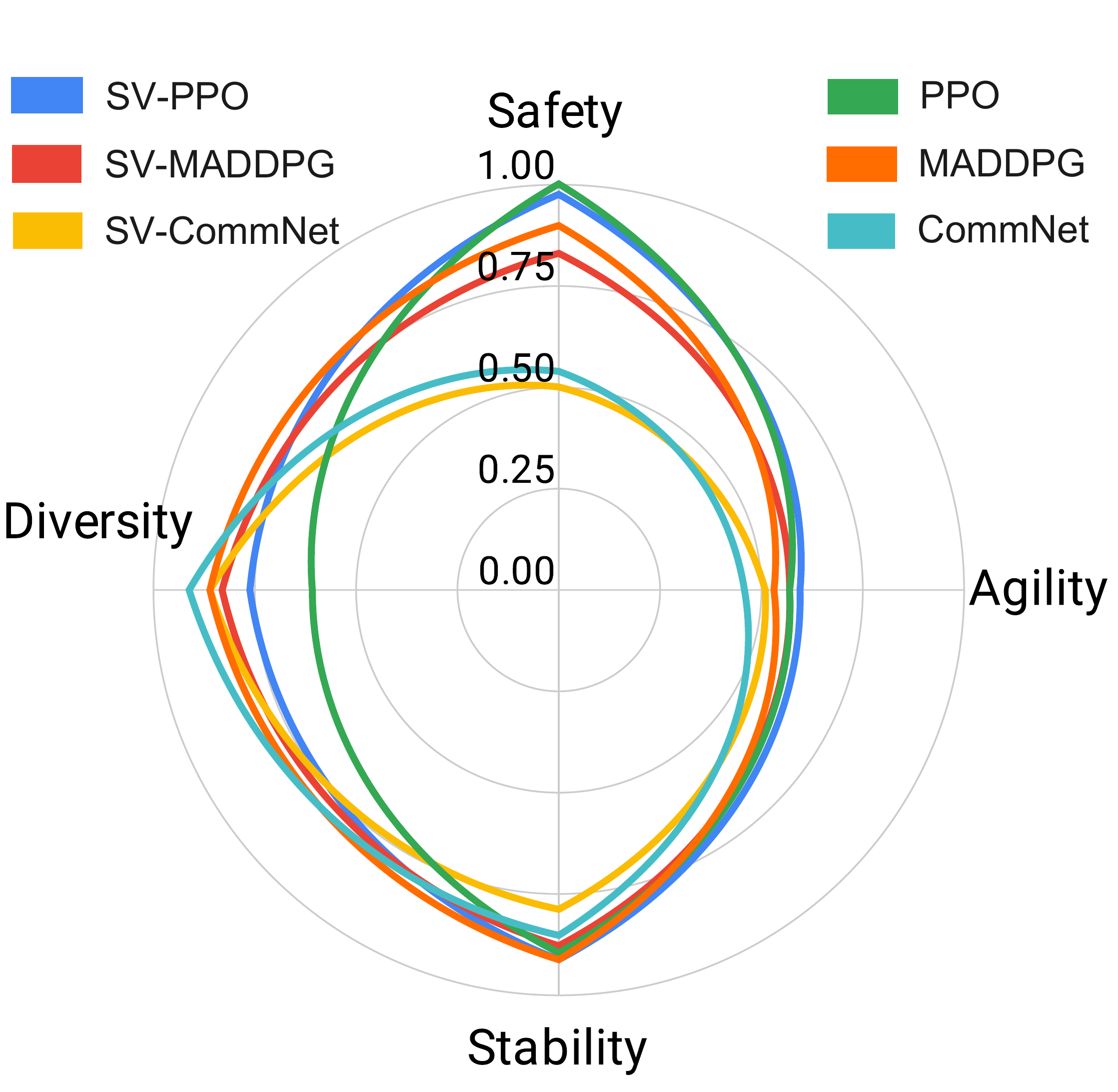}
    		\subcaption{Double Merge}
    	\end{subfigure}
    	\begin{subfigure}[b]{0.32\textwidth}
    		\centering
    		\includegraphics[width=\textwidth]{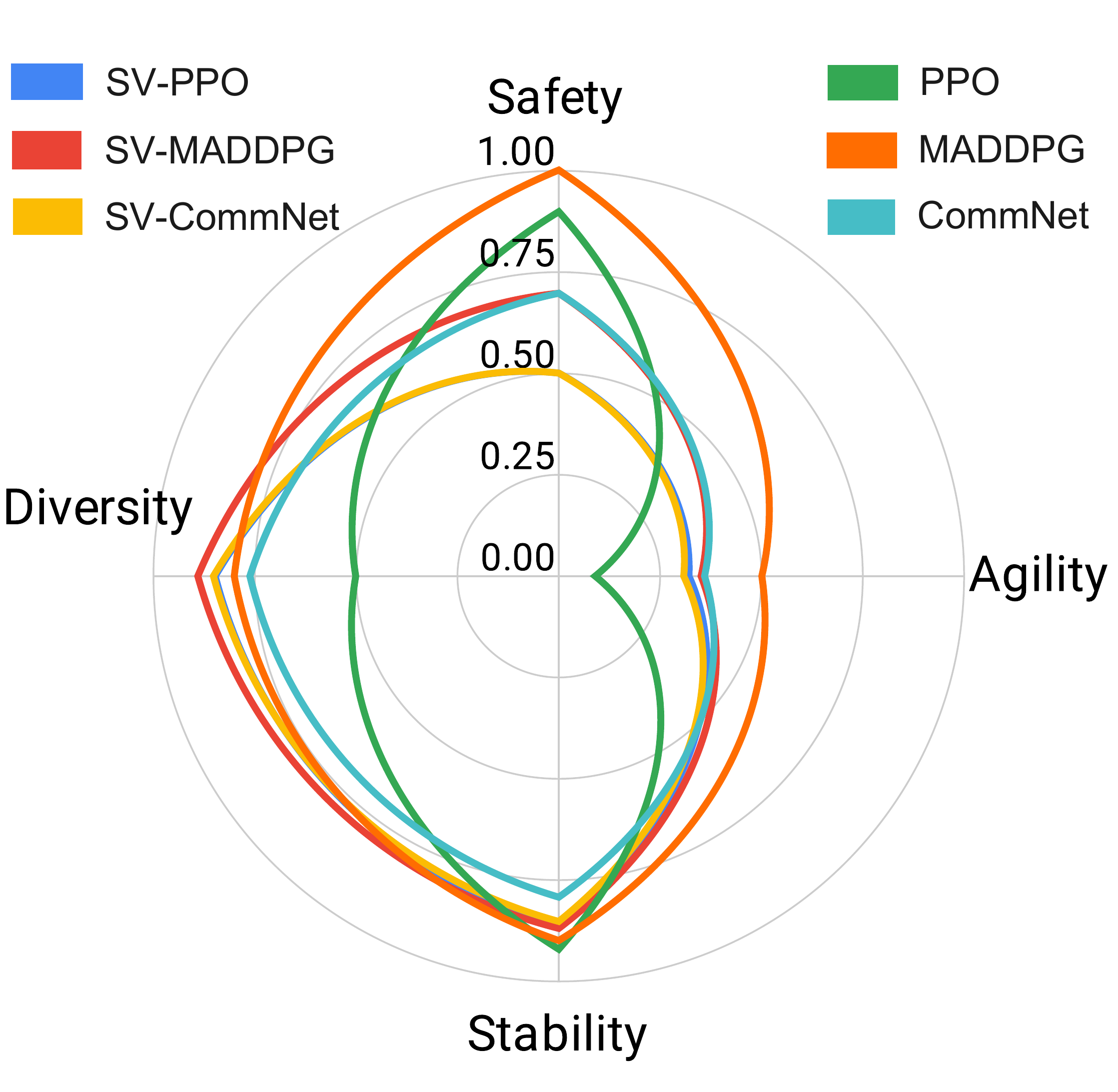}
    		\subcaption{Intersection}
    	\end{subfigure}
    	\caption{Results on behavior metrics. The larger the coverage, the more desirable the behavior. The wider scattered the curves, the more diverse the behaviors. ``SV-'' represents the algorithms interacting with social vehicles. See Figure~\ref{fig:radar-all} for additional plots.}
    	\label{fig:radar}
    \end{figure}

    \paragraph*{Behavior Analysis.} Figure~\ref{fig:radar} gives radar plots of four \textit{behavior metrics}: Safety, Agility, Stability, and (Control) Diversity. These plots allow us to analyze behavioral difference of the algorithms in different scenarios. In the Two-Way scenario, behavioral difference concentrates on Diversity. In Double-Merge and Intersection, behavioral difference covers both Diversity and Safety. In Intersection, Agility also sees a fair amount of behavior difference. The overall trend here is that as the difficulty of the scenarios increases, interaction among traffic participants become more frequent and more complex, which forces the agents to perform more complex behavior and makes it harder for algorithms to score higher on some metrics compared to the easier scenarios. Change in Agility between Two-Way and Intersection is a case in point: as the scenario gets harder, max achieved agility drops and the spread of agility becomes wider, across the algorithms.
    
%% file: conclusion.tex
\section{Conclusion}

We introduced \platform{}, an open-source platform that is dedicated to bringing together scalable multi-agent learning and scalable simulation of realistic driving interaction. Through our experiments and behavior analysis, we hope to show how the \platform{} platform as a whole can support investigations into how various MARL algorithms perform in the AD context. We hope this in turn can guide the development of new algorithms. Since its first internal release, \platform{} has successfully supported three autonomous driving competitions internationally,\footnote{SMARTS-supported competitions can be found at \url{https://drive-ml.com/}.} wherein thousands of submissions of agent models have been automatically evaluated for the leaderboards. We believe the value of \platform{} will grow in the coming years, as it exposes the realistic driving interaction challenge to more researchers and enables them to further their research through trying to solve this challenge in a principle way.

%% file: appendices.tex
\appendix

\section{Survey of AD Related Simulators}
\label{appendix:sims}

Table \ref{table:simulators} categorizes simulators relevant to autonomous driving research and explains how they may be related to \platform{}. As is clear from the table, simulators in the AD and MA\slash RL categories are more similar to \platform{} in terms of either their focus on autonomous driving or their emphasis on interactive behavior. Moreover, among the AD and MA\slash RL simulators in Table \ref{table:simulators}, we can discern two approaches to simulation of interactive behavior: \textit{data-replay simulators} and \textit{interactive simulators}.

\begin{table}[ht]
\begin{center}
\small
\begin{tabular}{| m{1.6cm} | m{2.4cm} | m{4.2cm} | m{4.2cm} |}
    \hline
    \textbf{Category} & \textbf{Simulator} & \textbf{Purpose} & \textbf{Comment} \\
    \hline
    \multirow{4}{*}{Automotive}     & CarMaker\cite{CarMaker91:online} & \multirow{4}{4.2cm}{Testing detailed designs regarding dynamics, safety, and performance of the individual physical vehicle.} & \multirow{4}{4.2cm}{Not concerned with interaction with other road users. Possible use as vehicle provider for \platform{}.} \\
    \cline{2-2}
                                    & CarSim \cite{Mechanic22:online} & & \\
    \cline{2-2}
                                    & Pro-SiVIC \cite{ProSiVIC51:online} & & \\
    \cline{2-2}
                                    & PreScan \cite{PreScanT59:online} & & \\
    \hline
    \multirow{4}{*}{Robotics}       & 4DV-Sim \cite{Automoti31:online} & \multirow{4}{4.2cm}{General robotics simulation for full sensor-control loop with realistic models.} & \multirow{4}{4.2cm}{With suitable robot models, could provide \platform{} with varieties of road users beyond just cars and trucks.} \\
    \cline{2-2}
                                    & CoppeliaSim \cite{Robotsim70:online} & & \\
    \cline{2-2}
                                    & Gazebo \cite{Gazebo10:online} & & \\
    \cline{2-2}
                                    & Webots \cite{cyberbot50:online} & & \\
    \hline
    \multirow{2}{*}{Games}          & GTA V \cite{GrandThe18:online} & An action-adventure game involving driving. & Its vivid graphics inspired its use for generating AD training data.  \\
    \cline{2-4}
                                    & TORCS \cite{torcs2013} & An open-source racing simulator. & Useful for testing RL algorithms but not AD due to racing focus. \\
    \hline
    \multirow{5}{*}{Traffic}        & Aimsun Next \cite{AimsunNe63:online} & \multirow{5}{4.2cm}{Traffic simulation at the \textit{microscopic} level, with motion and interaction of individual vehicles simulated.} & \multirow{5}{4.2cm}{Suitable for simulation of traffic flow but not interaction behavior. \platform{} currently uses SUMO as background traffic provider.} \\
    \cline{2-2}
                                    & MATSIM \cite{MATSimor56:online}  & & \\
    \cline{2-2}
                                    & MITSIMLab \cite{MITSIMLa86:online} & & \\
    \cline{2-2}
                                    & SUMO \cite{krajzewicz2002sumo} & & \\
    \cline{2-2}
                                    & VISSIM \cite{TrafficS34:online} & & \\
    \hline
    \multirow{5}{*}{AD}             & AirSim \cite{airsim2017fsr} & \multirow{5}{4.2cm}{Dedicated AD simulators. Limited interaction through scripting or data-replay.} & \multirow{5}{4.2cm}{Focus is on testing ego vehicle AI, not on solving interaction. Could use \platform{} to supply high-quality interaction.} \\
    \cline{2-2}
                                    & Apollo \cite{Apollo9:online} & & \\
    \cline{2-2}
                                    & CARLA \cite{dosovitskiy2017carla} & & \\
    \cline{2-2}
                                    & rFpro \cite{DrivingS93:online} & & \\
    \cline{2-2}
                                    & VTD \cite{VirtualT67:online} & & \\
    \hline
    \multirow{10}{*}{MA/RL}           & AIM4 \cite{AIMAuton91:online} & Multi-agent framework for managing autonomous vehicles at intersections. & Focused on intersection only and assuming all vehicles are autonomous. \\
    \cline{2-4}
                                    & BARK \cite{bernhard2020bark} & Multi-agent envs with extensible social agent models. & Similar to \platform{} in emphasizing importance of social agents; no explicit support for multi-agent research. \\
    \cline{2-4}
                                    & CarRacing \cite{brockman2016openai} & An RL env in OpenAI Gym. & Car racing for testing RL. \\
    \cline{2-4}
                                    & CityFlow \cite{cityflow} & Streamlined simulator for traffic optimization with RL. & Could provide background traffic in low-fidelity setting. \\
    \cline{2-4}
                                    & Duckietown \cite{duckieto16:online} & AD simulator for education. & For \href{https://www.duckietown.org/}{the Duckietown project}. \\
    \cline{2-4}
                                    & Flow \cite{flow-berkeley} & Microscopic traffic simulation for deep RL. & Wraps SUMO and Aimsun Next. \\
    \cline{2-4}
                                    & Gym-TORCS \cite{ugonamak4:online} & Gym-style env for TORCS & Shows need for standard envs. \\
    \cline{2-4}
                                    & highway-env \cite{highway-env} & Hand-coded interaction envs. & Shows need for interaction envs. \\
    \cline{2-4}
                                    & MACAD-Gym \cite{palanisamy2019multiagent} & Multi-agent connected car simulation using CARLA. & Links V2V communication with multi-agent coordination. \\
    \cline{2-4}
                                    & MADRaS \cite{madrassi35:online} & TORCS multi-agent wrapper. & Limited to racing. \\
    \hline
\end{tabular}
\end{center}
\caption{Overview of Existing Autonomous Driving Related Simulators.} 
\label{table:simulators}
\end{table}

Data-replay simulators are extensively used in the industry for real-world autonomous driving R\&D. Most simulated miles run by autonomous driving companies are based on replay of recorded data, against which the behavior of a new version of the ego vehicle AI may be assessed \cite{InsideWa70}. Because the recorded data capture how other road users actually reacted to the ego vehicle as well as to each other in real world trials, there is no question about the realism of their behavior. However, because the behavior of the new version of the ego vehicle AI is expected to diverge under at least some identical situations from the old version, in order to adequately assess the new version, the simulated road users must be able to reasonably react to the new behavior in a way that is not already reflected in the recorded data. This in turn requires intelligent agents that can control the simulated road users, especially the social vehicles \cite{TheDisen76, DragoAng38}. However, technical specifics about how to add interactive intelligence to data-replay simulators are virtually never published.

Interactive simulators, instead of relying on recorded data, try to build the necessary interactive behaviors from ground up by providing intelligent behavior models for the simulated road users. CARLA \cite{dosovitskiy2017carla}, highway-env \cite{highway-env}, BARK \cite{bernhard2020bark}, and even SUMO \cite{krajzewicz2002sumo} are such examples. In the fully multi-agent case (e.g. AIM4 \cite{AIMAuton91:online}), there may not even be any real difference between social vehicle AI and ego vehicle AI. To date, however, the behavior of road users in these simulators is typically controlled by simple rules, scripts, or highly limited models. This results in overly simplistic and rigid behavior that falls far short of the complexity and diversity of real interactions on the road.

\platform{} is expected to be a platform that bootstraps realistic and diverse behavior models (Figure \ref{figure:bootstrap}) that can be used to \textit{make data-replay simulators more interactive} and \textit{make interactive simulators more realistic}.

\section{SMARTS Simulations}
\label{appendix:smarts-demos}
To give an intuitive sense of what \platform{} is like when up and running, we have included several screenshots from a few sample simulation runs.

\begin{figure}[h]
  \centering
  \includegraphics[width=0.9\textwidth]{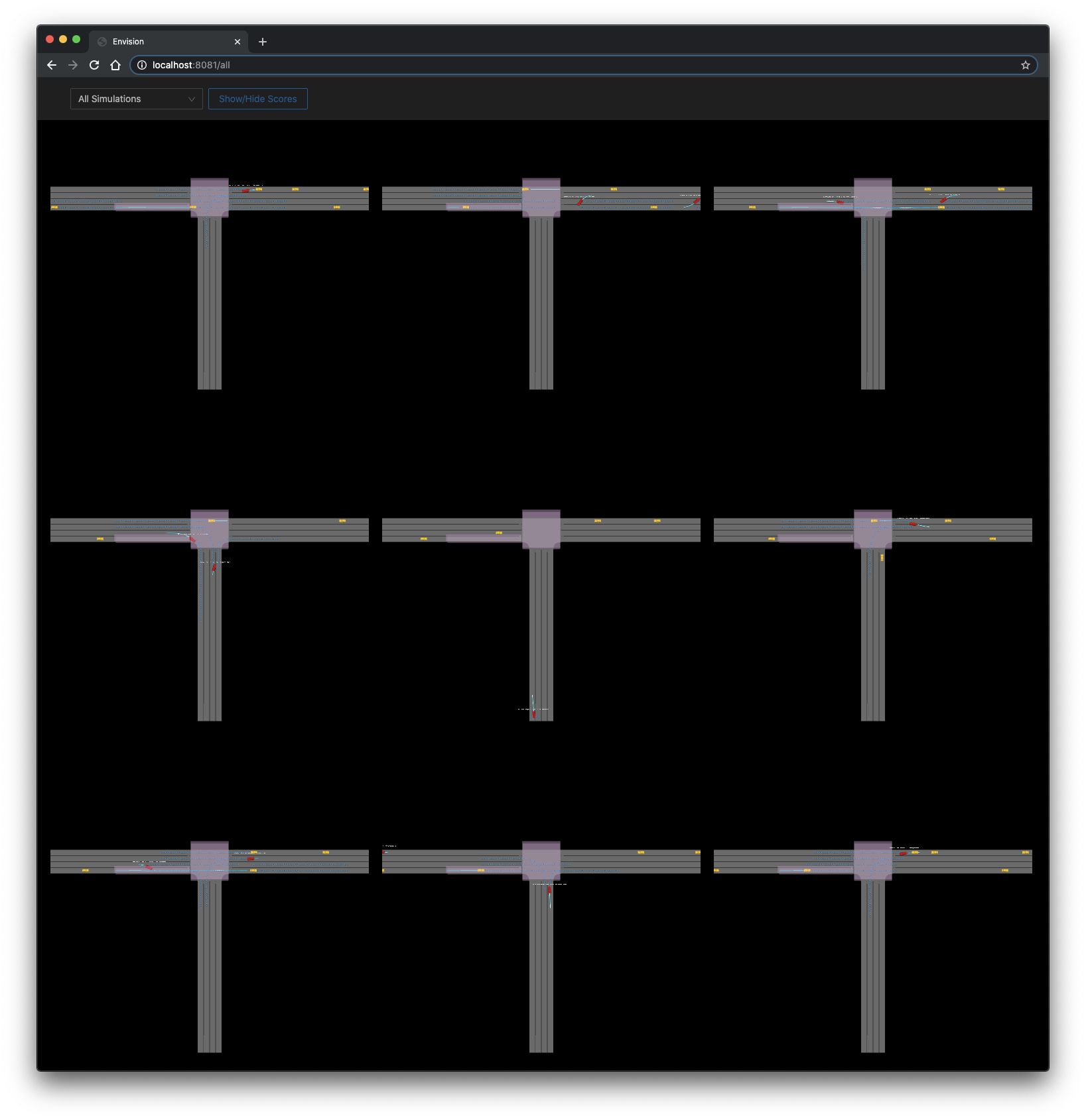}
  \caption{Multi-Instance and Web-Base Visualization. The \platform{} platform supports multiple instances running simultaneously. Instances are distributed across multiple processes or across networked machines. To collect diverse interaction data quickly, each instance can be configured to run its own set of multiple scenarios by sampling through them. In this example, 9 concurrent instances are simultaneously visualized in the browser.}
  \label{figure:multi-instance}
\end{figure}

\begin{figure}[h]
  \centering
  \includegraphics[width=0.9\textwidth]{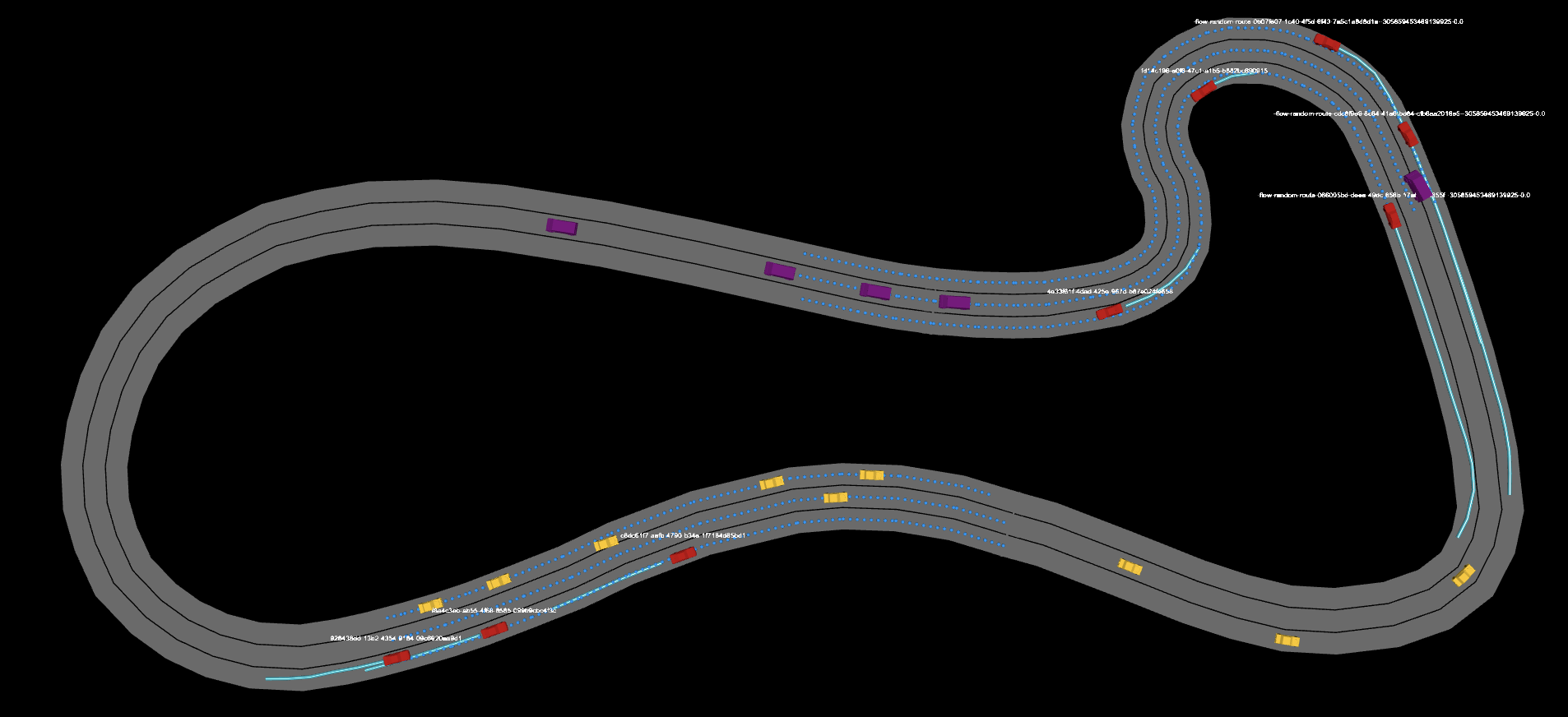}
  \caption{Multi-Agent. Multiple ego agents in training (red vehicles) can run simultaneously within a shared SMARTS instance. Each agent runs in a separate process and can also run remotely.}
  \label{figure:multi-agent}
  \vspace{-1mm}
\end{figure}

\begin{figure}[h]
  \centering
  \begin{subfigure}[b]{0.44\textwidth}
    \includegraphics[width=\textwidth]{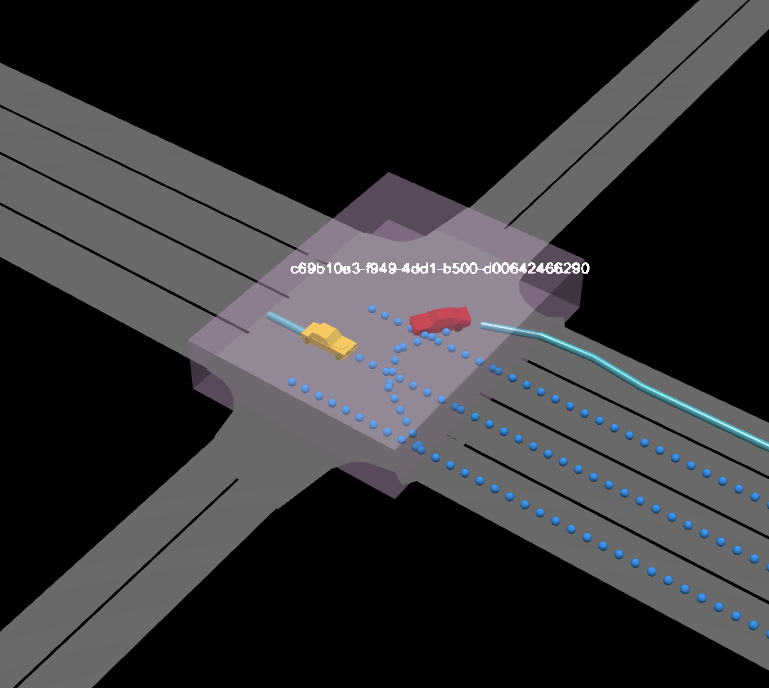}
    \caption{Bubble}
    \label{fig:bubble}
  \end{subfigure}
  \hspace{1mm}
  \begin{subfigure}[b]{0.44\textwidth}
    \includegraphics[width=\textwidth]{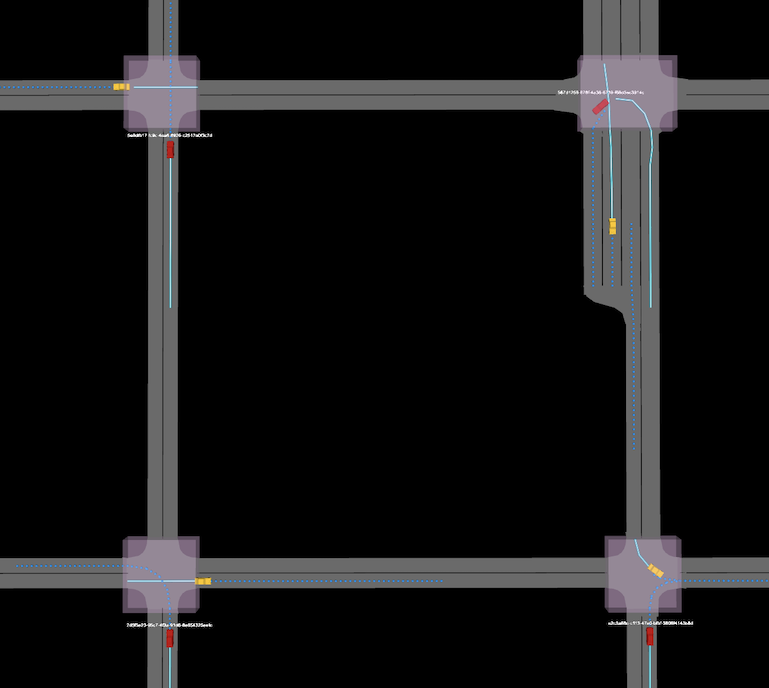}
    \caption{Multiple Bubbles}
    \label{fig:bubbles}
  \end{subfigure}
  \caption{Bubbles. Bubbles are regions in which social vehicles (yellow) may be controlled by agents from the Social Agent Zoo and interact meaningfully with ego vehicles (red).}
  \vspace{-1mm}
\end{figure}

\section{Controllers and Action Spaces}\label{appendix:controllers}
In \platform{}, agent's action space depends on the controller. Table~\ref{tab:controllers} shows the mapping between controllers and action spaces.

\begin{table}
\renewcommand\arraystretch{1.0}
\begin{center}
\begin{tabular}{|c|c|c|}
\hline
\textbf{Controller}                        & \textbf{Action Space Type}  & \textbf{Control Command Dimensions}       \\ \hline
\multirow{2}{*}{LaneFollowingController}   & \multirow{2}{*}{Mixed}      & target speed                      \\ \cline{3-3} 
                                           &                             & lane change (+1 or 0 or -1)            \\ \hline
TrajectoryTrackingController               & -                           & trajectory                        \\ \hline
\multirow{3}{*}{ActuatorDynamicController} & \multirow{3}{*}{Continuous} & throttle                          \\ \cline{3-3} 
                                           &                             & brake                             \\ \cline{3-3} 
                                           &                             & steering rate in rad \\ \hline
\multirow{3}{*}{ContinuousController}      & \multirow{3}{*}{Continuous} & throttle                          \\ \cline{3-3} 
                                           &                             & brake                             \\ \cline{3-3} 
                                           &                             & steering                          \\ \hline
\end{tabular}
\end{center}
\caption{Mapping from controllers to action spaces.}
\label{tab:controllers}
\end{table}

\section{Algorithm Library} \label{appendix:algorithms}
\begin{figure}
  \centering
  \includegraphics[width=0.95\textwidth]{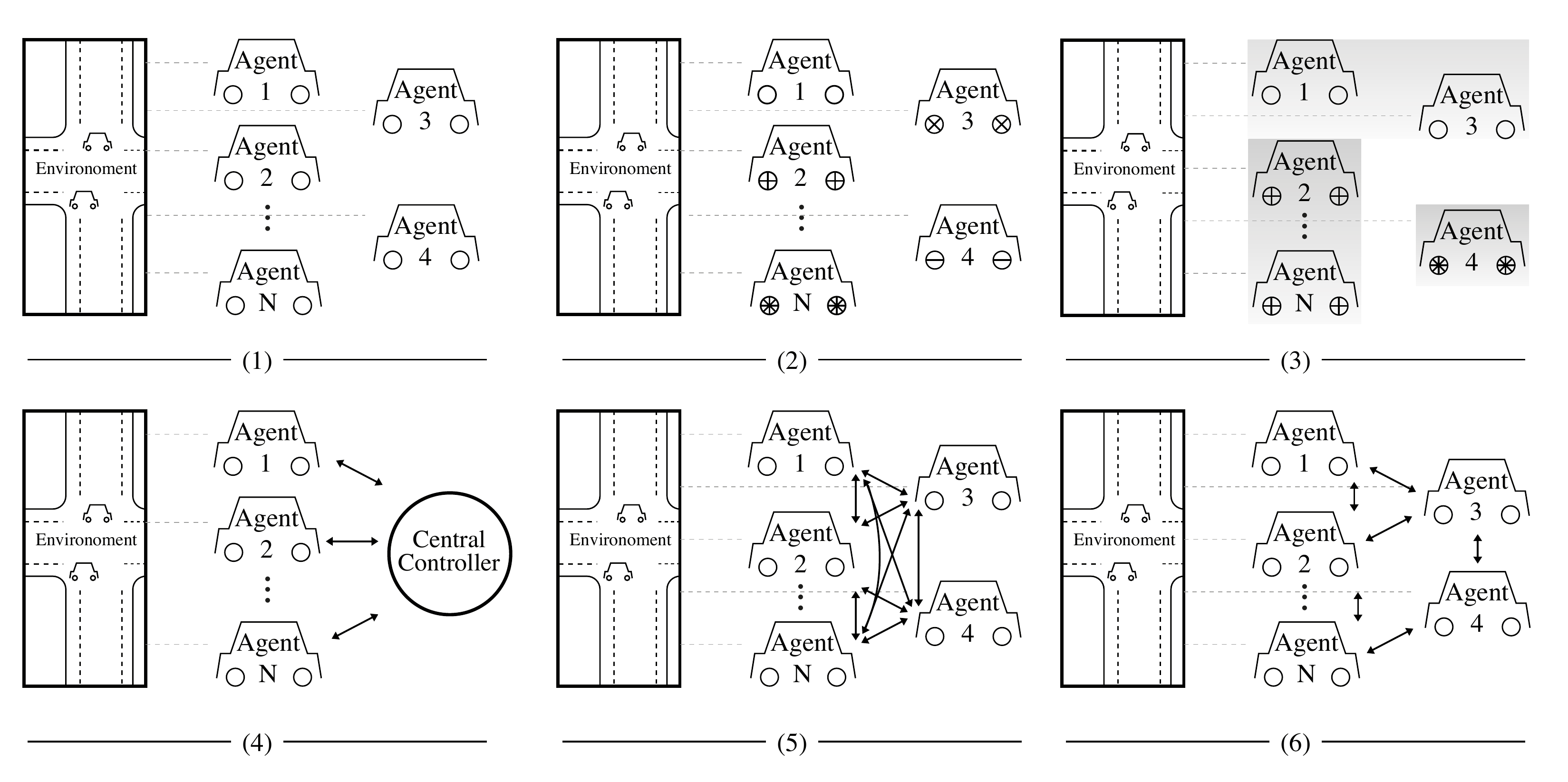}
  \caption{MARL Paradigms. The \textit{fully decentralized paradigm} has three sub-settings: (1) parameter sharing, (2) non-parameter-sharing, and (3) group sharing, respectively. (4) is the \textit{fully centralized training paradigm}, i.e., a central controller controls all agents to make decisions. (5) is the \textit{CTDE paradigm}, wherein each agent shares its information with other agents globally, but makes decisions independently. (6) represents the \textit{networked agent learning paradigm}, which differs from CTDE in that each agent shares information only with its networked neighbors.}
  \label{figure:marl-paradigm}
\end{figure}

We have integrated a large set of popular MARL algorithms. Table~\ref{tab:algorithms} gives an overview of these algorithms and tags them according to 1) paradigm, 2) communication or not, and 3) framework dependency. The four paradigms are fully centralized training, fully decentralized, centralized training \& decentralized execution (CTDE) and networked agent learning. They are illustrated in Figure~\ref{figure:marl-paradigm}. 

\begin{table}
\renewcommand\arraystretch{1.4}
\begin{center}
    \resizebox{\columnwidth}{!}{
    \begin{tabular}{|c|c|c|c|}
    \hline
    \textbf{Paradigm } & \textbf{Algorithm} & \textbf{Communication} & \textbf{Framework} \\ \hline
                                                 & BiCNet \cite{peng2017multiagent} $\star$ & Yes & malib \\ \cline{2-4} 
    \multirow{-2}{*}{Fully centralized training} & CommNet\cite{sukhbaatar2016learning} $\star$ & Yes & RLlib/malib \\ \hline
                                                 & Indepedent Q \cite{tan1993multi} & No & RLlib \\ \cline{2-4} 
                                                 & Independent PG \cite{sutton1999policy} & No & RLlib \\ \cline{2-4} 
                                                 & Independent AC \cite{mnih2016asynchronous} & No & RLlib/malib \\ \cline{2-4} 
                                                 & PR2 \cite{wen2018probabilistic} & No & malib \\ \cline{2-4} 
                                                 & ROMMEO \cite{tian2019regularized} & No & malib \\ \cline{2-4} 
    \multirow{-6}{*}{Fully decentralized}        & Supervised Opponent Modeling & No & malib \\ \hline
                                                 & Centralized V & No & RLlib \\ \cline{2-4}
                                                 & MAAC $\star$ & No & RLlib \\ \cline{2-4}
                                                 & MADDPG \cite{lowe2017multi} & No & RLlib/malib \\ \cline{2-4} 
                                                 & MF-AC/Q \cite{yang2018mean} $\star$ & No & RLlib/malib \\ \cline{2-4} 
                                                 & COMA \cite{comafoerster} & No & PyMARL \\ \cline{2-4} 
                                                 & VDN \cite{sunehag2018value} & No & PyMARL \\ \cline{2-4} 
                                                 & QMIX \cite{rashid2018qmix} & No & PyMARL \\ \cline{2-4} 
                                                 & QTRAN \cite{son2019qtran} & No & PyMARL \\ \cline{2-4} 
                                                 & MAVEN \cite{mahajan2019maven} & No & PyMARL \\ \cline{2-4} 
    \multirow{-10}{*}{CTDE} & Q-DPP \cite{osogami2019determinantal} $\star$        & No & PyMARL \\ \hline
    Networked agent learning                     & Networked Fitted-Q \cite{zhang2018finite} $\star$ & graph & RLlib \\ \hline
    \end{tabular}}
\end{center}
\caption{Algorithms available in \platform{}. Those tagged with $\star$ are our own implementations.} %\yaodong{@ming, i think it will be a good idea to integrate this table with Fig. \ref{figure:murl-paradigm}.}
\label{tab:algorithms}
\end{table}
%\textbf{d}

% \begin{table}[h]
%     \renewcommand\arraystretch{1.3}
%     \begin{center}
%     \resizebox{\columnwidth}{!}{
%         \begin{tabular}{ccccccc}
%         \hline
%         \multirow{2}{*}{Scenario} & \multicolumn{3}{c}{No Social Vehicle} & \multicolumn{3}{c}{Random Social Vehicle} \\
%                  & MF-AC & Net-Q & CommNet & MF-AC & Net-Q & CommNet \\ \hline
%         Two-Way & 0/0.8 & 0/0.3 & \textbf{0/0.96} & 0.45/0.5 & 0.4/0.2 & \textbf{0.25/0.65} \\ 
%         Double Merge & 0.6/0.4 & 0.7/0.25 & \textbf{0.46/0.45} & 0.7/0.3 & 0.8/0.2 & \textbf{0.5/0.5} \\
%         Intersection & 0.54/0.4  & 0.4/0.23 & \textbf{0.3/0.7} & 0.62/0.37 & 0.75/0.2 & \textbf{0.5/0.45}  \\ \hline
%         \end{tabular}
%     }
%     \end{center}
%     \caption{Average collision rate / completion rate of MF-AC, Networked Fitted-Q and CommNet. Not fine tuned.}
%     \label{tab:cc-suplementary}
% \end{table}

\section{Experiment Results}

\begin{figure}[ht]
	\centering
	\begin{subfigure}[b]{.45\textwidth}
		\centering
		\includegraphics[width=\textwidth]{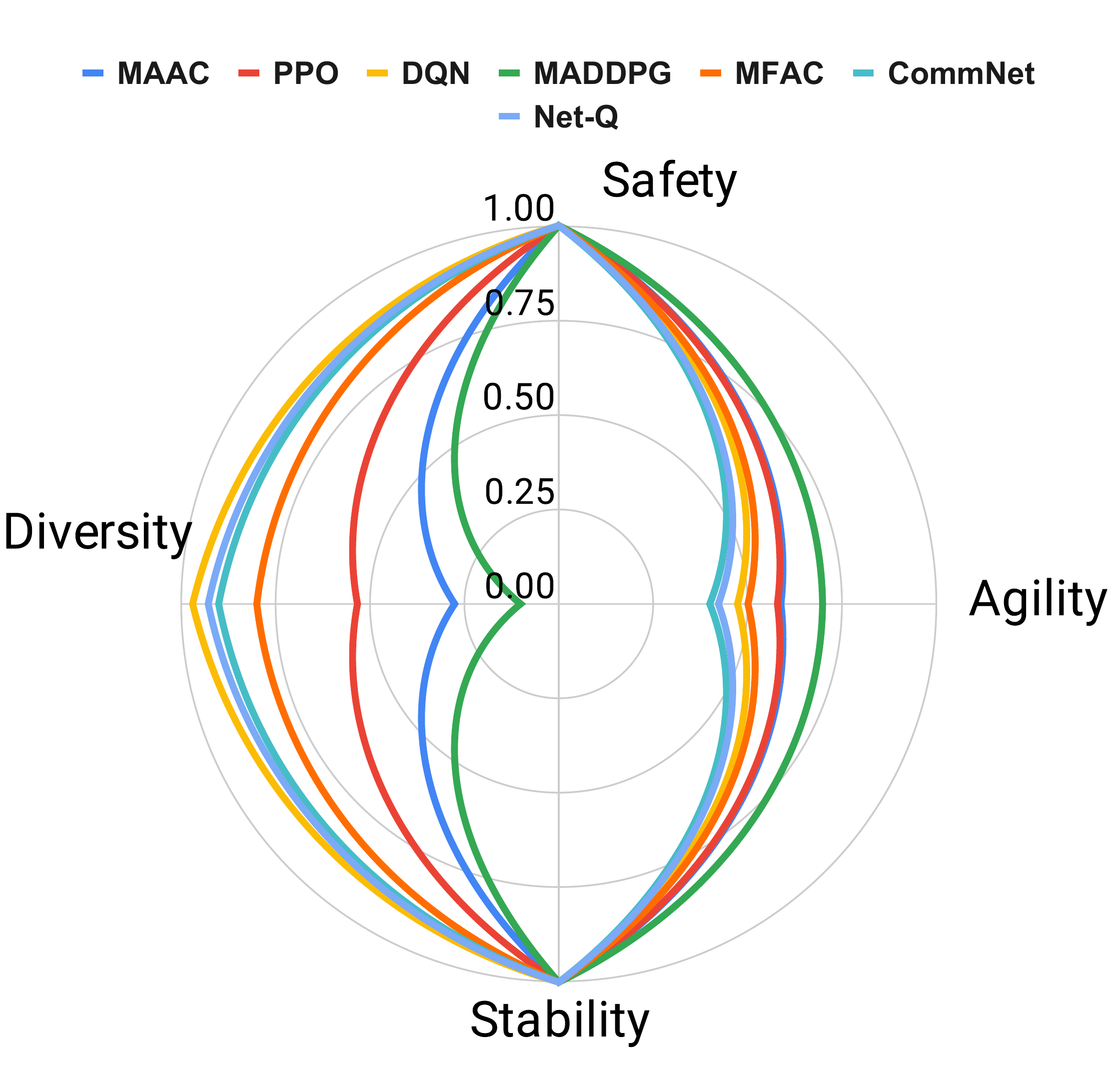}
		\subcaption{Two-Way: No SV}
	\end{subfigure}
	\begin{subfigure}[b]{.443\textwidth}
	    \setlength{\abovecaptionskip}{0.3cm} 
		\centering
		\includegraphics[width=\textwidth]{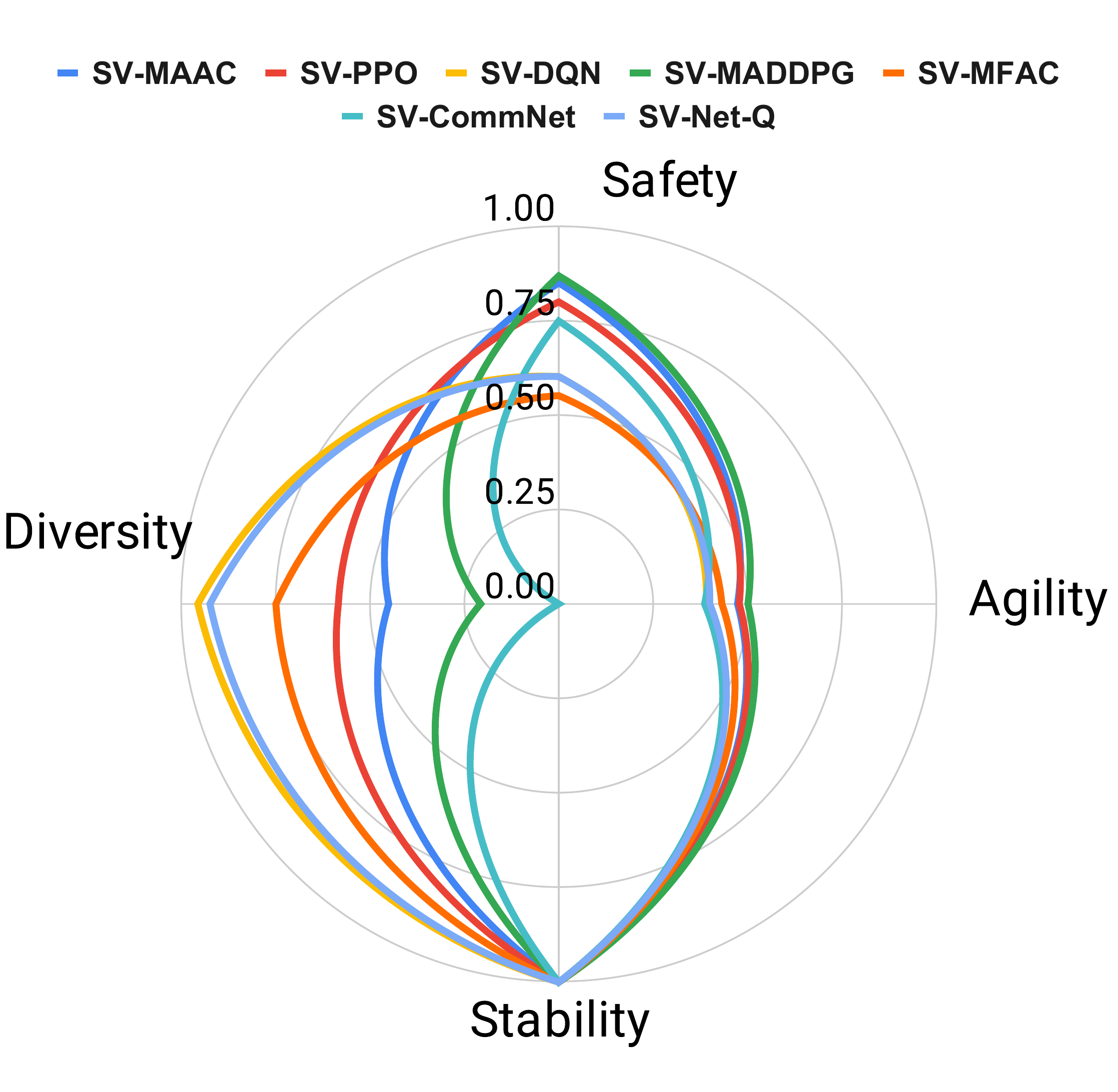}
		\subcaption{Two-Way: SV}
	\end{subfigure}
	\begin{subfigure}[b]{.45\textwidth}
	    \setlength{\abovecaptionskip}{0.3cm} 
		\centering
		\includegraphics[width=\textwidth]{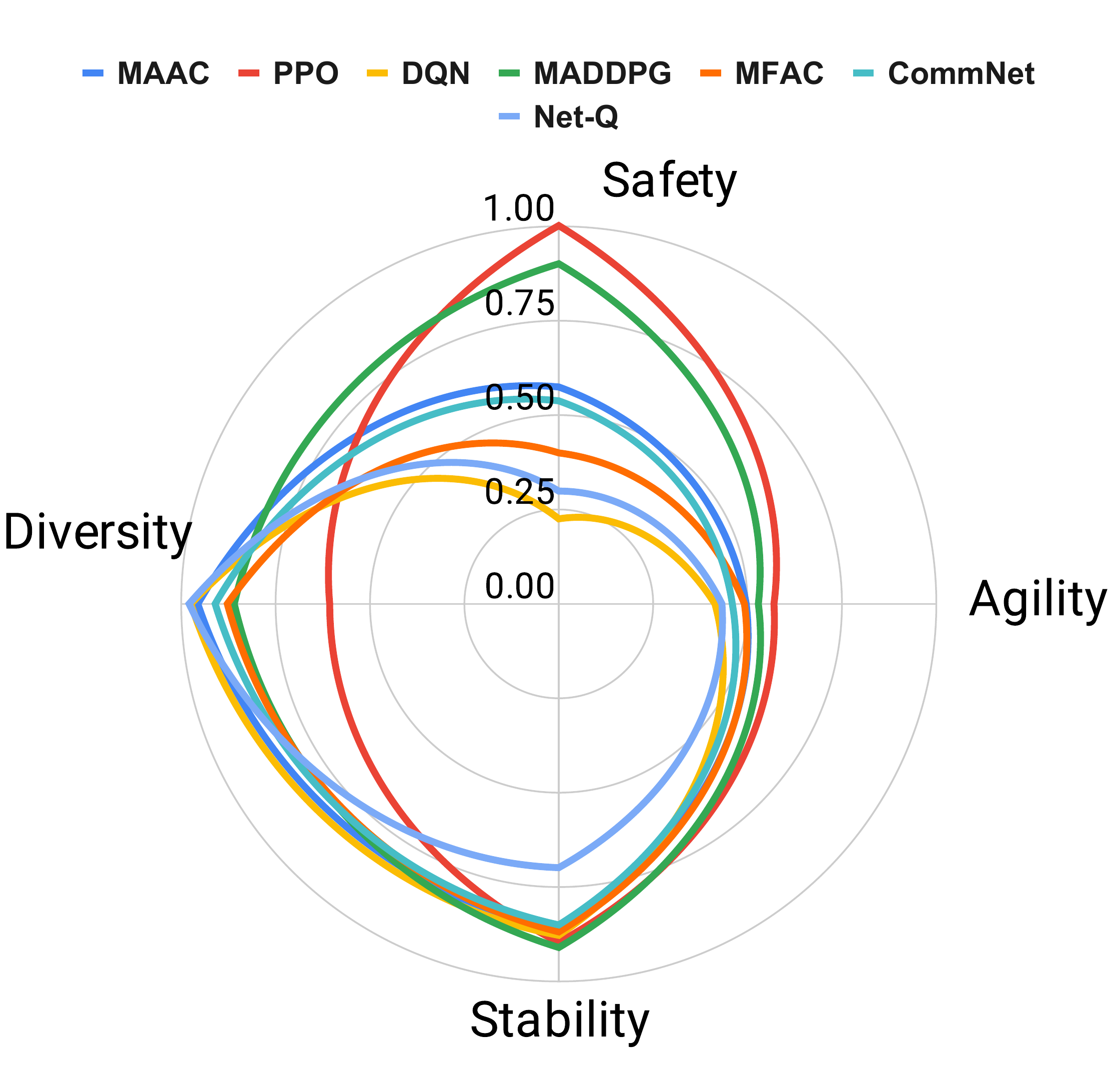}
		\subcaption{Double Merge: No SV}
	\end{subfigure}
	\begin{subfigure}[b]{.45\textwidth}
	    \setlength{\abovecaptionskip}{0.3cm} 
		\centering
		\includegraphics[width=\textwidth]{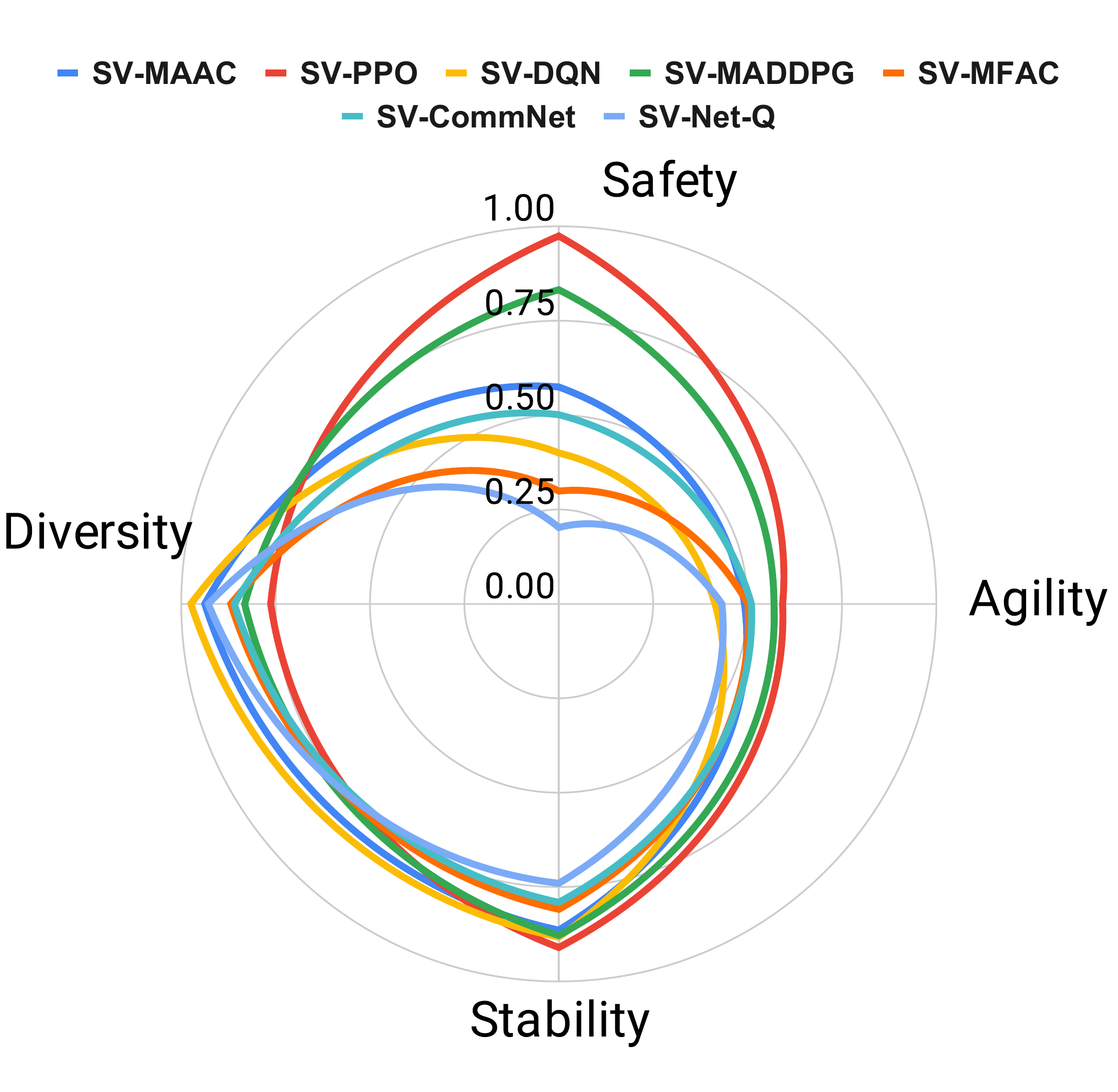}
		\subcaption{Double Merge: SV}
	\end{subfigure}
	\begin{subfigure}[b]{.45\textwidth}
	    \setlength{\abovecaptionskip}{0.3cm} 
		\centering
		\includegraphics[width=\textwidth]{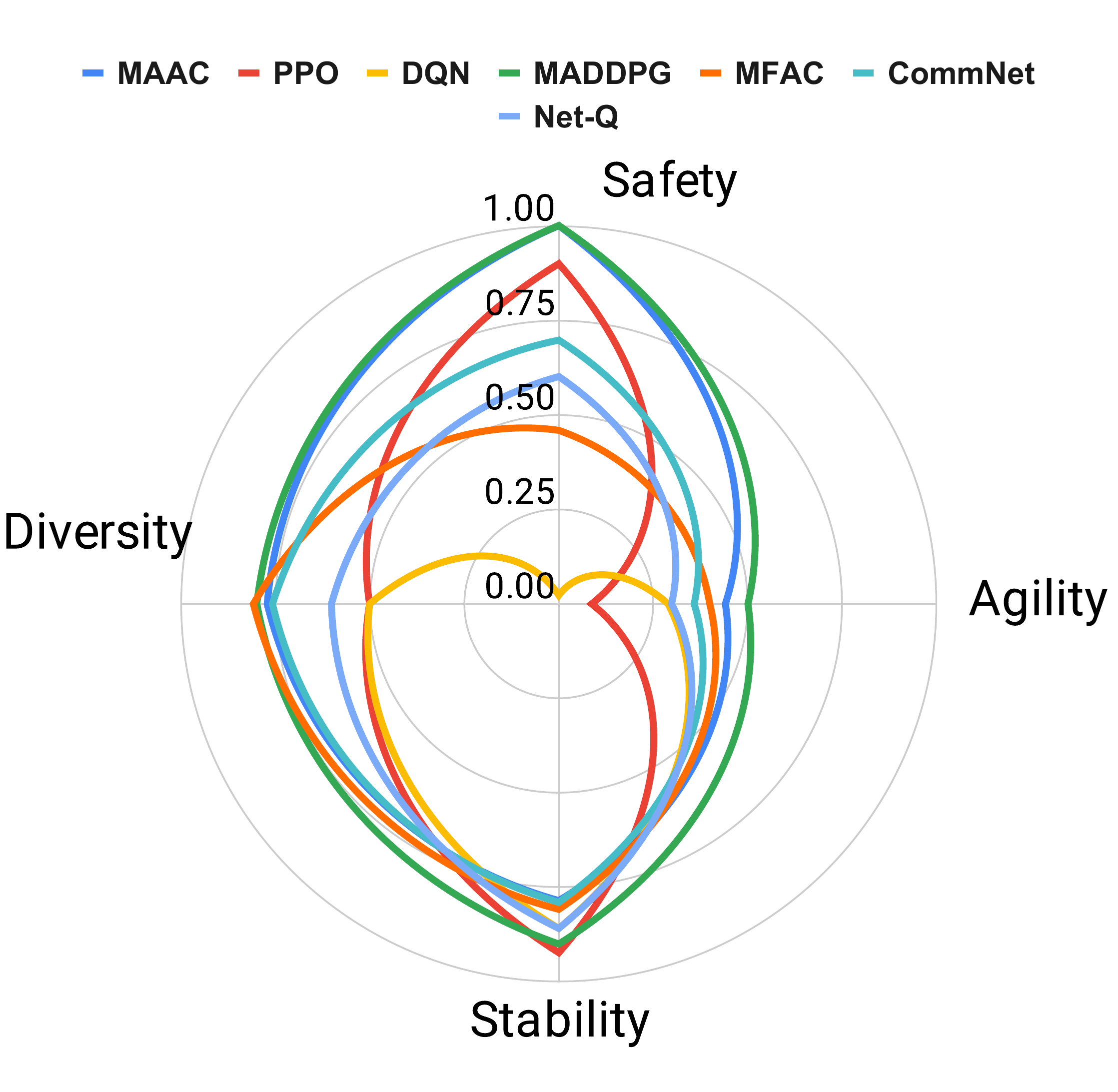}
		\subcaption{Intersection: No SV}
	\end{subfigure}
	\begin{subfigure}[b]{.45\textwidth}
	    \setlength{\abovecaptionskip}{0.3cm} 
		\centering
		\includegraphics[width=\textwidth]{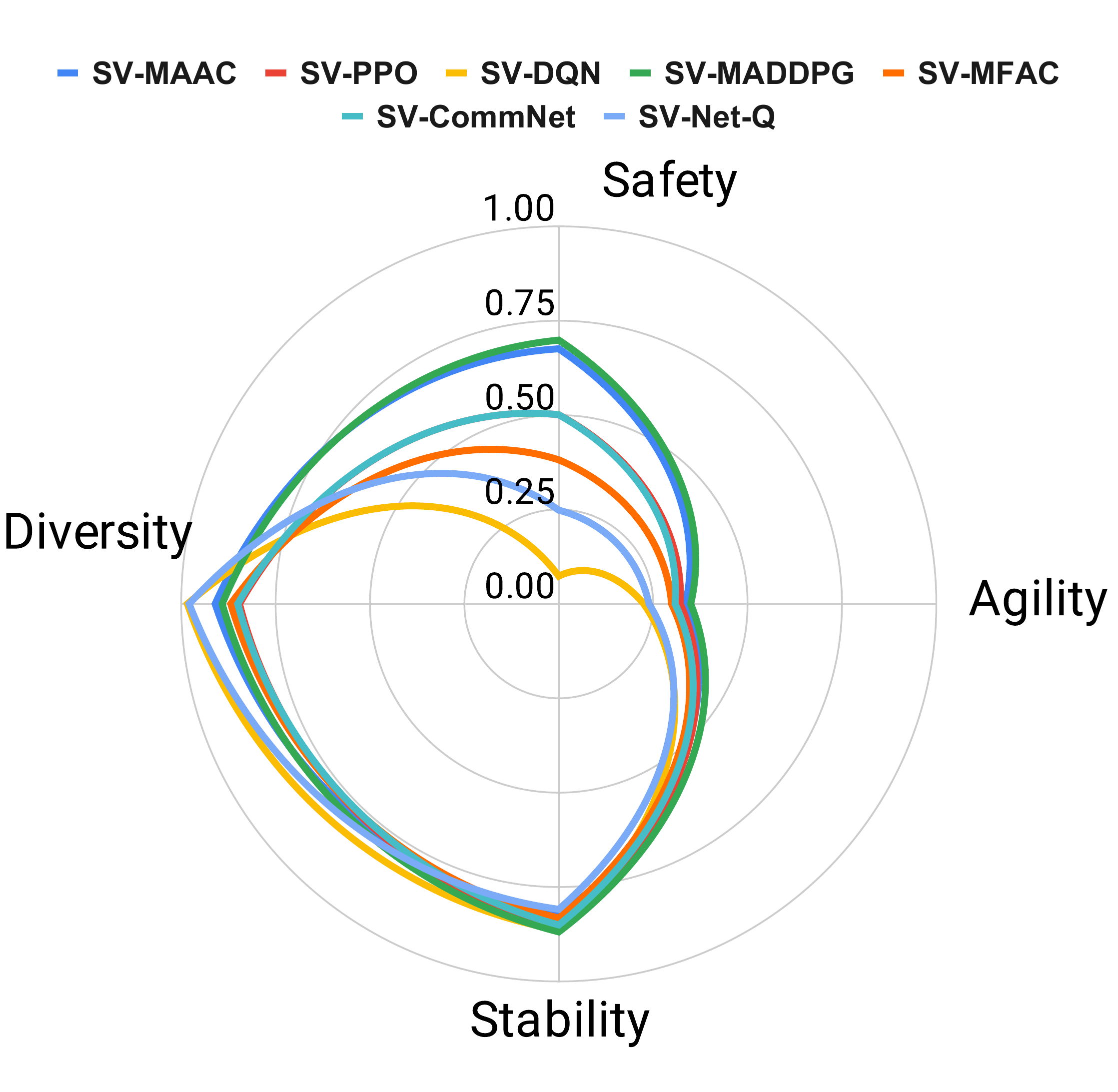}
		\subcaption{Intersection: SV}
	\end{subfigure}
% 	\begin{subfigure}[b]{0.32\textwidth}
% 		\centering
% 		\includegraphics[width=\textwidth]{imgs/scenarios/intersection1.pdf}
% 		\subcaption{Intersection}
% 	\end{subfigure}
	\caption{Results on behavior metrics. The larger the coverage, the more desirable the behavior. The wider scattered the curves, the more diverse the behaviors. ``SV'' represents the algorithms interacting with social vehicles. ``No SV'' means no social vehicles.}
	\label{fig:radar-all}
\end{figure}

Table \ref{tab:cc-big} and Figure \ref{fig:radar-all} provide additional details about our experiment results.

\begin{table}
    \renewcommand\arraystretch{1.2}
    \begin{center}
    \resizebox{\columnwidth}{!}{
        \begin{tabular}{ccccccc}
        \toprule
        \multirow{2}{*}{Algorithm} & \multicolumn{3}{c}{Scenario - No Social Vehicle} & \multicolumn{3}{c}{Scenario - Random Social Vehicle} \\
                                   & Two-Way     & Double Merge     & Intersection    & Two-Way        & Double Merge     & Intersection     \\ \midrule
        DQN                        & 0/0.97      & 0.77/0.23        & 0.83/0.20       & 0.40/0.60      & 0.60/0.23        & 0.92/0.05        \\
        PPO                        & \textbf{0/1}         & \textbf{0/1}              & 0.1/0.07        & 0.25/0.75      & \textbf{0.02/0.98}        & 0.50/0.45        \\
        MAAC                       & \textbf{0/1}         & 0.42/0.58        & \textbf{0/1}             & 0.25/0.75      & 0.42/0.6         & 0.32/0.68        \\
        MFAC                       & 0/0.8       & 0.6/0.4          & 0.54/0.4        & 0.45/0.5       & 0.7/0.3          & 0.62/0.37        \\
        Net-Q                      & 0/0.3       & 0.7/0.25         & 0.4/0.23        & 0.4/0.2        & 0.8/0.2          & 0.75/0.2         \\
        CommNet                    & 0/0.96      & 0.46/0.45        & 0.3/0.7         & 0.25/0.65      & 0.5/0.5          & 0.5/0.45         \\
        MADDPG                     & \textbf{0/1}         & 0.1/0.9          & \textbf{0/1}             & \textbf{0.13/0.87}      & 0.17/0.8         & \textbf{0.30/0.7}         \\ \bottomrule
        \end{tabular}
    }
    \end{center}
    \caption{Average collision rate / completion rate of selected baselines. Net-Q: Networked Fitted-Q.}
    \label{tab:cc-big}
\end{table}

\section{User Examples}

\subsection{Scenario Generation}
In SMARTS, a scenario is composed of map, traffic flow, agent missions, and optionally but importantly bubble specification. Given the map, users can easily create scenarios with specific requirements by writing a scenario configuration file. An example of scenario configuration and generation is given below.

\begin{verbatim}
# Configuration: in `scenario.py'
# 1. agent mission initialization and generation
agent_missions = [
    Mission(Route(begin=("top_left", 0, 10), end=("down_right", 0, 30))),
    ...
]

gen_missions(scenario=scenario, missions=agent_missions)

# 2. define social vehicle with diverse behavior
impatient_car = TrafficActor(
    name="car",
    speed=Distribution(sigma=0.2, mean=1.0),
    lane_changing_model=LaneChangingModel(impatience=1, cooperative=0.25),
    junction_model=JunctionModel(
        drive_after_red_time=1.5, 
        drive_after_yellow_time=1.0, 
        impatience=1.0
    ),
)

patient_car = TrafficActor(
    name="car",
    speed=Distribution(sigma=0.2, mean=0.8),
    lane_changing_model=LaneChangingModel(impatience=0, cooperative=0.5),
    junction_model=JunctionModel(drive_after_yellow_time=1.0, impatience=0.5)
)

# 3. traffic flow customization
traffic = Traffic(
    flows=[
        Flow(
            route=Route(begin=("down_left", 0, 30), end=("merge", 0, 150),),
            rate=1.,
            actors={impatient_car: 0.5, patient_car: 0.5},
        ),
        ...
    ]
)

gen_traffic(scenario, traffic, name="example")

# one command to do scenario generation
make build-scenario scenario=${SCENARIO_DIR}
\end{verbatim}

\subsection{Building Agents}
\platform{} provide 10 built-in agent types that cover frequently used combinations of observation-sensor and action-controller combinations. These are available through the agent interface. The combination of agent interface, agent policy, and a set of adapters form the agent spec. Concrete agent instances are built according to the agent spec.

\begin{verbatim}
# AgentSpec for agent customization
class AgentSpec:
    ...
    policy_builder: Callable[..., AgentPolicy] = None
    policy_params: Optional[Any] = None
    observation_adapter: Callable = lambda obs: obs
    action_adapter: Callable = lambda act: act
    reward_adapter: Callable = lambda obs, reward: reward
    info_adapter: Callable = lambda obs, reward, info: info
    ...

# use case: build agents
agent_spec = AgentSpec(
    interface=AgentInterface.from_type(AgentType.Laner),
    policy_params={"policy_function": lambda _: "keep_lane"},
    policy_builder=AgentPolicy.from_function,
)

env = gym.make(
    "smarts.env:hiway-v0",
    scenarios=["scenarios/loop"],
    agent_specs={agent_id: agent_spec},
)

agent = agent_spec.build_agent()
\end{verbatim}

\subsection{Running Single-agent \& Multi-agent Training}
In \platform{}, it is easy to build a single-agent or multi-agent training experiment. The key steps include agent initialization, algorithm specification, and training scenario specification. Multiple training scenarios can be automatically loaded and used for multi-task learning. 

\begin{verbatim}
# initializing multiple agents
agents = {
    agent_id: AgentSpec(
        **config.agent, interface=AgentInterface(**config.interface)
    )
    for agent_id in agent_ids
}

# initializing environment with one or more scenarios
env_config = {
    "seed": 42,
    "scenarios": [str(scenario_path), ...],  # specifying scenarios
    "headless": args.headless,
    "agent_specs": agents,
}

policies = {k: config.policy for k in agents}

tune_config = {
    ...
    "env": RLlibHiWayEnv,
    "env_config": env_config,
    "multiagent": {
        "policies": policies,
        "policy_mapping_fn": lambda agent_id: agent_id,
    },
    ...
}

# run experiment
analysis = tune.run(
    ...
    config=tune_config,
)
\end{verbatim}
\vspace{-15pt}
\subsection{Running Evaluation}
We implemented a benchmarking runner to do algorithm evaluation. The benchmarking runner's \texttt{Episode} recorder and \texttt{Metric} class allow us to record the steps and perform evaluation accordingly.

\begin{verbatim}
@dataclass
class Episode(EpisodeLog):
    def record_step(**kwargs):
        ...

class Metric:
    def log_step(self, observations, actions, rewards, dones,
                    infos, episode):
        ...

    def compute(self):
        """Evaluate algorithm based on records with the given metrics."""
        ...
        
metrics_obj = Metric(num_episodes=10)
while not done and step < num_steps:
    # ...
    metrics_obj.log_step(observations, actions, rewards, 
                    dones, infos, episode=episode)
    # ...
_ = metrics_obj.compute()
\end{verbatim}

%% file: main.bbl
\begin{thebibliography}{83}
\providecommand{\natexlab}[1]{#1}
\providecommand{\url}[1]{\texttt{#1}}
\expandafter\ifx\csname urlstyle\endcsname\relax
  \providecommand{\doi}[1]{doi: #1}\else
  \providecommand{\doi}{doi: \begingroup \urlstyle{rm}\Url}\fi

\bibitem[Buehler et~al.(2007)Buehler, Iagnemma, and Singh]{buehler20072005}
M.~Buehler, K.~Iagnemma, and S.~Singh.
\newblock \emph{The 2005 DARPA grand challenge: the great robot race},
  volume~36.
\newblock Springer, 2007.

\bibitem[Waymo(2020)]{waymo2020safety}
Waymo.
\newblock Waymo safety report.
\newblock \url{https://bit.ly/2T4vRl4}, 09 2020.
\newblock (Accessed on 09/22/2020).

\bibitem[Badue et~al.(2020)Badue, Guidolini, Carneiro, Azevedo, Cardoso,
  Forechi, Jesus, Berriel, Paix{\~a}o, Mutz, et~al.]{badue2020self}
C.~Badue, R.~Guidolini, R.~V. Carneiro, P.~Azevedo, V.~B. Cardoso, A.~Forechi,
  L.~Jesus, R.~Berriel, T.~M. Paix{\~a}o, F.~Mutz, et~al.
\newblock Self-driving cars: A survey.
\newblock \emph{Expert Systems with Applications}, page 113816, 2020.

\bibitem[Stewart()]{stewart}
J.~Stewart.
\newblock Humans just can't stop rear-ending self-driving cars—let's figure
  out why.
\newblock \url{https://bit.ly/3jfcfFs}.
\newblock (Accessed on 09/23/2020).

\bibitem[Efrati()]{WaymoB72}
A.~Efrati.
\newblock Waymo’s big ambitions slowed by tech trouble.
\newblock \url{https://bit.ly/31lKwgt}.
\newblock (Accessed on 09/22/2020).

\bibitem[Siddiqui()]{Someofth6}
F.~Siddiqui.
\newblock Some of the biggest critics of {Waymo} and other self-driving cars
  are the silicon valley residents who know how they work.
\newblock \url{https://wapo.st/30UAX6R}.
\newblock (Accessed on 09/22/2020).

\bibitem[Efrati({\natexlab{a}})]{WaymoRid1}
A.~Efrati.
\newblock Waymo riders describe experiences on the road, {\natexlab{a}}.
\newblock URL
  \url{https://www.theinformation.com/articles/waymo-riders-describe-experiences-on-the-road}.
\newblock (Accessed on 09/22/2020).

\bibitem[Efrati({\natexlab{b}})]{WaymoB80}
A.~Efrati.
\newblock Waymo’s backseat drivers: Confidential data reveals self-driving
  taxi hurdles, {\natexlab{b}}.
\newblock URL
  \url{https://www.theinformation.com/articles/waymos-backseat-drivers-confidential-data-reveals-self-driving-taxi-hurdles}.
\newblock (Accessed on 09/22/2020).

\bibitem[International(2016)]{sae2018levels}
S.~International.
\newblock Taxonomy and definitions for terms related to driving automation
  systems for on-road motor vehicles.
\newblock 2016.

\bibitem[Roughgarden(2005)]{roughgarden2005selfish}
T.~Roughgarden.
\newblock \emph{Selfish routing and the price of anarchy}, volume 174.
\newblock MIT press Cambridge, 2005.

\bibitem[Kendall et~al.(2019)Kendall, Hawke, Janz, Mazur, Reda, Allen, Lam,
  Bewley, and Shah]{kendall2019learning}
A.~Kendall, J.~Hawke, D.~Janz, P.~Mazur, D.~Reda, J.-M. Allen, V.-D. Lam,
  A.~Bewley, and A.~Shah.
\newblock Learning to drive in a day.
\newblock In \emph{ICRA}, pages 8248--8254. IEEE, 2019.

\bibitem[Shalev-Shwartz et~al.(2016)Shalev-Shwartz, Shammah, and
  Shashua]{mobileye2016}
S.~Shalev-Shwartz, S.~Shammah, and A.~Shashua.
\newblock Safe, multi-agent, reinforcement learning for autonomous driving.
\newblock 10 2016.

\bibitem[Schwarting et~al.(2019)Schwarting, Pierson, Alonso-Mora, Karaman, and
  Rus]{schwarting2019}
W.~Schwarting, A.~Pierson, J.~Alonso-Mora, S.~Karaman, and D.~Rus.
\newblock Social behavior for autonomous vehicles.
\newblock \emph{PNAS}, 116\penalty0 (50):\penalty0 24972--24978, 2019.
\newblock ISSN 0027-8424.
\newblock \doi{10.1073/pnas.1820676116}.
\newblock URL \url{https://www.pnas.org/content/116/50/24972}.

\bibitem[Wang et~al.(2020)Wang, Everett, and How]{wang2020r}
R.~E. Wang, M.~Everett, and J.~P. How.
\newblock R-{MADDPG} for partially observable environments and limited
  communication.
\newblock \emph{arXiv preprint arXiv:2002.06684}, 2020.

\bibitem[AIM()]{AIMAuton91:online}
{AIM}: Autonomous intersection management.
\newblock \url{http://www.cs.utexas.edu/~aim/}.
\newblock (Accessed on 09/22/2020).

\bibitem[Zhang et~al.(2019{\natexlab{a}})Zhang, Chen, Huang, Li, Yang, Zhang,
  and Wang]{zhang2019bi}
H.~Zhang, W.~Chen, Z.~Huang, M.~Li, Y.~Yang, W.~Zhang, and J.~Wang.
\newblock Bi-level actor-critic for multi-agent coordination.
\newblock \emph{arXiv preprint arXiv:1909.03510}, 2019{\natexlab{a}}.

\bibitem[Zhang et~al.(2019{\natexlab{b}})Zhang, Feng, Liu, Ding, Zhu, Zhou,
  Zhang, Yu, Jin, and Li]{cityflow}
H.~Zhang, S.~Feng, C.~Liu, Y.~Ding, Y.~Zhu, Z.~Zhou, W.~Zhang, Y.~Yu, H.~Jin,
  and Z.~Li.
\newblock {C}ity{F}low: A multi-agent reinforcement learning environment for
  large scale city traffic scenario.
\newblock In \emph{WWW}, page 3620–3624. ACM, 2019{\natexlab{b}}.

\bibitem[Wu et~al.(2017)Wu, Kreidieh, Parvate, Vinitsky, and
  Bayen]{flow-berkeley}
C.~Wu, A.~Kreidieh, K.~Parvate, E.~Vinitsky, and A.~M. Bayen.
\newblock Flow: Architecture and benchmarking for reinforcement learning in
  traffic control.
\newblock \emph{ArXiv}, abs/1710.05465, 2017.

\bibitem[Wei et~al.(2019)Wei, Xu, Zhang, Zheng, Zang, Chen, Zhang, Zhu, Xu, and
  Li]{wei2019colight}
H.~Wei, N.~Xu, H.~Zhang, G.~Zheng, X.~Zang, C.~Chen, W.~Zhang, Y.~Zhu, K.~Xu,
  and Z.~Li.
\newblock Colight: Learning network-level cooperation for traffic signal
  control.
\newblock In \emph{CIKM}, pages 1913--1922, 2019.

\bibitem[Krajzewicz et~al.(2002)Krajzewicz, Hertkorn, R{\"o}ssel, and
  Wagner]{krajzewicz2002sumo}
D.~Krajzewicz, G.~Hertkorn, C.~R{\"o}ssel, and P.~Wagner.
\newblock {SUMO} ({S}imulation of {U}rban {MO}bility)---an open-source traffic
  simulation.
\newblock In \emph{MESM}, pages 183--187, 2002.

\bibitem[{Dresner} and {Stone}(2004)]{dresner2004aim}
K.~{Dresner} and P.~{Stone}.
\newblock Multiagent traffic management: a reservation-based intersection
  control mechanism.
\newblock In \emph{AAMAS 2004.}, pages 530--537, 2004.

\bibitem[Barcel\'{\i} et~al.(2005)Barcel\'{\i}, Codina, Casas, Ferrer, and
  Garc\'{\i}a]{barcel2005micro}
J.~Barcel\'{\i}, E.~Codina, J.~Casas, J.~L. Ferrer, and D.~Garc\'{\i}a.
\newblock Microscopic traffic simulation: A tool for the design, analysis and
  evaluation of intelligent transport systems.
\newblock \emph{J. Intell. Robotics Syst.}, 41\penalty0 (2–3):\penalty0
  173–203, Jan. 2005.
\newblock ISSN 0921-0296.
\newblock \doi{10.1007/s10846-005-3808-2}.
\newblock URL \url{https://doi.org/10.1007/s10846-005-3808-2}.

\bibitem[Dosovitskiy et~al.(2017)Dosovitskiy, Ros, Codevilla, Lopez, and
  Koltun]{dosovitskiy2017carla}
A.~Dosovitskiy, G.~Ros, F.~Codevilla, A.~Lopez, and V.~Koltun.
\newblock Carla: An open urban driving simulator.
\newblock \emph{arXiv preprint arXiv:1711.03938}, 2017.

\bibitem[Leurent(2018)]{highway-env}
E.~Leurent.
\newblock An environment for autonomous driving decision-making.
\newblock \url{https://github.com/eleurent/highway-env}, 2018.

\bibitem[Bernhard et~al.(2020)Bernhard, Esterle, Hart, and
  Kessler]{bernhard2020bark}
J.~Bernhard, K.~Esterle, P.~Hart, and T.~Kessler.
\newblock {BARK}: Open behavior benchmarking in multi-agent environments.
\newblock \emph{arXiv preprint arXiv:2003.02604}, 2020.

\bibitem[Todorov et~al.(2012)Todorov, Erez, and Tassa]{todorov2012mujoco}
E.~Todorov, T.~Erez, and Y.~Tassa.
\newblock {MuJoCo}: A physics engine for model-based control.
\newblock In \emph{2012 IEEE/RSJ International Conference on Intelligent Robots
  and Systems}, pages 5026--5033. IEEE, 2012.

\bibitem[Vinyals et~al.(2017)Vinyals, Ewalds, Bartunov, Georgiev, Vezhnevets,
  Yeo, Makhzani, K{\"u}ttler, Agapiou, Schrittwieser,
  et~al.]{vinyals2017starcraft}
O.~Vinyals, T.~Ewalds, S.~Bartunov, P.~Georgiev, A.~S. Vezhnevets, M.~Yeo,
  A.~Makhzani, H.~K{\"u}ttler, J.~Agapiou, J.~Schrittwieser, et~al.
\newblock {StarCraft II}: A new challenge for reinforcement learning.
\newblock \emph{arXiv preprint arXiv:1708.04782}, 2017.

\bibitem[Silver et~al.(2017)Silver, Hubert, Schrittwieser, Antonoglou, Lai,
  Guez, Lanctot, Sifre, Kumaran, Graepel, et~al.]{silver2017mastering2}
D.~Silver, T.~Hubert, J.~Schrittwieser, I.~Antonoglou, M.~Lai, A.~Guez,
  M.~Lanctot, L.~Sifre, D.~Kumaran, T.~Graepel, et~al.
\newblock Mastering chess and shogi by self-play with a general reinforcement
  learning algorithm.
\newblock \emph{arXiv preprint arXiv:1712.01815}, 2017.

\bibitem[Vinyals et~al.(2019)Vinyals, Babuschkin, Czarnecki, Mathieu, Dudzik,
  Chung, Choi, Powell, Ewalds, Georgiev,
  et~al.]{vinyals2019alphastargrandmaster}
O.~Vinyals, I.~Babuschkin, W.~M. Czarnecki, M.~Mathieu, A.~Dudzik, J.~Chung,
  D.~H. Choi, R.~Powell, T.~Ewalds, P.~Georgiev, et~al.
\newblock Grandmaster level in {StarCraft II} using multi-agent reinforcement
  learning.
\newblock \emph{Nature}, 575\penalty0 (7782):\penalty0 350--354, 2019.

\bibitem[Moritz et~al.(2018)Moritz, Nishihara, Wang, Tumanov, Liaw, Liang,
  Elibol, Yang, Paul, Jordan, et~al.]{moritz2018ray}
P.~Moritz, R.~Nishihara, S.~Wang, A.~Tumanov, R.~Liaw, E.~Liang, M.~Elibol,
  Z.~Yang, W.~Paul, M.~I. Jordan, et~al.
\newblock Ray: A distributed framework for emerging {AI} applications.
\newblock In \emph{OSDI}, pages 561--577, 2018.

\bibitem[Samvelyan et~al.(2019)Samvelyan, Rashid, de~Witt, Farquhar, Nardelli,
  Rudner, Hung, Torr, Foerster, and Whiteson]{samvelyan2019starcraft}
M.~Samvelyan, T.~Rashid, C.~S. de~Witt, G.~Farquhar, N.~Nardelli, T.~G. Rudner,
  C.-M. Hung, P.~H. Torr, J.~Foerster, and S.~Whiteson.
\newblock The {StarCraft} multi-agent challenge.
\newblock \emph{arXiv preprint arXiv:1902.04043}, 2019.

\bibitem[Wen()]{yingwenm12:online}
Y.~Wen.
\newblock Malib: A multi-agent learning framework.
\newblock \url{https://github.com/ying-wen/malib}.
\newblock (Accessed on 09/23/2020).

\bibitem[Casas et~al.(2010)Casas, Ferrer, Garcia, Perarnau, and
  Torday]{casas2010traffic}
J.~Casas, J.~L. Ferrer, D.~Garcia, J.~Perarnau, and A.~Torday.
\newblock Traffic simulation with {Aimsun}.
\newblock In \emph{Fundamentals of traffic simulation}, pages 173--232.
  Springer, 2010.

\bibitem[Bul()]{BulletRe59}
Bullet real-time physics simulation.
\newblock \url{https://pybullet.org/wordpress/}.
\newblock (Accessed on 09/23/2020).

\bibitem[Scheiderer et~al.(2019)Scheiderer, Thun, and
  Meisen]{scheiderer2019bezier}
C.~Scheiderer, T.~Thun, and T.~Meisen.
\newblock B{\'e}zier curve based continuous and smooth motion planning for
  self-learning industrial robots.
\newblock \emph{Procedia Manufacturing}, 38:\penalty0 423--430, 2019.

\bibitem[Sopasakis et~al.(2020)Sopasakis, Fresk, and Patrinos]{open2020}
P.~Sopasakis, E.~Fresk, and P.~Patrinos.
\newblock {OpEn}: Code generation for embedded nonconvex optimization.
\newblock In \emph{IFAC World Congress}, Berlin, 2020.

\bibitem[Liang et~al.(2018)Liang, Liaw, Nishihara, Moritz, Fox, Goldberg,
  Gonzalez, Jordan, and Stoica]{liang2018rllib}
E.~Liang, R.~Liaw, R.~Nishihara, P.~Moritz, R.~Fox, K.~Goldberg, J.~Gonzalez,
  M.~Jordan, and I.~Stoica.
\newblock Rllib: Abstractions for distributed reinforcement learning.
\newblock In \emph{ICML}, pages 3053--3062, 2018.

\bibitem[Zhang et~al.(2018)Zhang, Yang, Liu, Zhang, and Basar]{zhang2018fully}
K.~Zhang, Z.~Yang, H.~Liu, T.~Zhang, and T.~Basar.
\newblock Fully decentralized multi-agent reinforcement learning with networked
  agents.
\newblock In \emph{ICML}, pages 5872--5881, 2018.

\bibitem[Mnih et~al.(2013)Mnih, Kavukcuoglu, Silver, Graves, Antonoglou,
  Wierstra, and Riedmiller]{mnih2013playing}
V.~Mnih, K.~Kavukcuoglu, D.~Silver, A.~Graves, I.~Antonoglou, D.~Wierstra, and
  M.~Riedmiller.
\newblock Playing atari with deep reinforcement learning.
\newblock \emph{arXiv preprint arXiv:1312.5602}, 2013.

\bibitem[Schulman et~al.(2017)Schulman, Wolski, Dhariwal, Radford, and
  Klimov]{schulman2017proximal}
J.~Schulman, F.~Wolski, P.~Dhariwal, A.~Radford, and O.~Klimov.
\newblock Proximal policy optimization algorithms.
\newblock \emph{arXiv preprint arXiv:1707.06347}, 2017.

\bibitem[Yang et~al.(2018)Yang, Luo, Li, Zhou, Zhang, and Wang]{yang2018mean}
Y.~Yang, R.~Luo, M.~Li, M.~Zhou, W.~Zhang, and J.~Wang.
\newblock Mean field multi-agent reinforcement learning.
\newblock In \emph{International Conference on Machine Learning}, pages
  5571--5580, 2018.

\bibitem[Lowe et~al.(2017)Lowe, Wu, Tamar, Harb, Abbeel, and
  Mordatch]{lowe2017multi}
R.~Lowe, Y.~Wu, A.~Tamar, J.~Harb, P.~Abbeel, and I.~Mordatch.
\newblock Multi-agent actor-critic for mixed cooperative-competitive
  environments.
\newblock \emph{arXiv preprint arXiv:1706.02275}, 2017.

\bibitem[Zhang et~al.(2018)Zhang, Yang, and Basar]{zhang2018networked}
K.~Zhang, Z.~Yang, and T.~Basar.
\newblock Networked multi-agent reinforcement learning in continuous spaces.
\newblock In \emph{2018 IEEE CDC}, pages 2771--2776. IEEE, 2018.

\bibitem[Sukhbaatar et~al.(2016)Sukhbaatar, Fergus,
  et~al.]{sukhbaatar2016learning}
S.~Sukhbaatar, R.~Fergus, et~al.
\newblock Learning multiagent communication with backpropagation.
\newblock In \emph{NIPS}, pages 2244--2252, 2016.

\bibitem[Car()]{CarMaker91:online}
Carmaker | {IPG} {A}utomotive.
\newblock
  \url{https://ipg-automotive.com/products-services/simulation-software/carmaker/}.
\newblock (Accessed on 09/22/2020).

\bibitem[Mec()]{Mechanic22:online}
Mechanical simulation.
\newblock \url{https://www.carsim.com/}.
\newblock (Accessed on 09/22/2020).

\bibitem[Pro()]{ProSiVIC51:online}
{P}ro-{S}i{VIC} 2017.0 | my{ESI}.
\newblock
  \url{https://myesi.esi-group.com/downloads/software-downloads/pro-sivic-2017.0}.
\newblock (Accessed on 09/22/2020).

\bibitem[Pre()]{PreScanT59:online}
{P}re{S}can | {TASS} {I}nternational.
\newblock \url{https://tass.plm.automation.siemens.com/prescan}.
\newblock (Accessed on 09/22/2020).

\bibitem[Aut()]{Automoti31:online}
Automotive-simulator – 4{D}-virtualiz.
\newblock \url{https://www.4d-virtualiz.com/en/automotive-simulator/}.
\newblock (Accessed on 09/22/2020).

\bibitem[Rob()]{Robotsim70:online}
Robot simulator coppeliasim: create, compose, simulate, any robot - coppelia
  robotics.
\newblock \url{https://www.coppeliarobotics.com/}.
\newblock (Accessed on 09/22/2020).

\bibitem[Gaz()]{Gazebo10:online}
Gazebo.
\newblock \url{http://gazebosim.org/}.
\newblock (Accessed on 09/22/2020).

\bibitem[cyb()]{cyberbot50:online}
Webots robot simulator.
\newblock \url{https://github.com/cyberbotics/webots}.
\newblock (Accessed on 09/22/2020).

\bibitem[{Rockstar Games}()]{GrandThe18:online}
{Rockstar Games}.
\newblock {Grand Theft Auto V}.
\newblock \url{https://www.rockstargames.com/games/V}.
\newblock (Accessed on 09/22/2020).

\bibitem[Loiacono et~al.(2013)Loiacono, Cardamone, and Lanzi]{torcs2013}
D.~Loiacono, L.~Cardamone, and P.~L. Lanzi.
\newblock Simulated car racing championship: Competition software manual, 2013.

\bibitem[Aimsun()]{AimsunNe63:online}
Aimsun.
\newblock {Aimsun Next}: Your personal mobility modeling lab.
\newblock \url{https://www.aimsun.com/aimsun-next/}.
\newblock (Accessed on 09/22/2020).

\bibitem[MAT()]{MATSimor56:online}
{MATSim}.org.
\newblock \url{https://www.matsim.org/}.
\newblock (Accessed on 09/22/2020).

\bibitem[MIT()]{MITSIMLa86:online}
{MITSIML}ab | {INTELLIGENT TRANSPORTATION SYSTEMS LAB}.
\newblock \url{https://its.mit.edu/software/mitsimlab}.
\newblock (Accessed on 09/22/2020).

\bibitem[{PTV Group}()]{TrafficS34:online}
{PTV Group}.
\newblock Traffic simulation software.
\newblock \url{https://www.ptvgroup.com/en/solutions/products/ptv-vissim/}.
\newblock (Accessed on 09/22/2020).

\bibitem[Shah et~al.(2017)Shah, Dey, Lovett, and Kapoor]{airsim2017fsr}
S.~Shah, D.~Dey, C.~Lovett, and A.~Kapoor.
\newblock {A}ir{S}im: {H}igh-fidelity visual and physical simulation for
  autonomous vehicles.
\newblock In \emph{FSR}, 2017.
\newblock URL \url{https://arxiv.org/abs/1705.05065}.

\bibitem[Apo()]{Apollo9:online}
Apollo.
\newblock \url{https://apollo.auto/platform/simulation.html}.
\newblock (Accessed on 09/22/2020).

\bibitem[Dri()]{DrivingS93:online}
Driving simulation for autonomous driving, {ADAS}, vehicle dynamics and
  motorsport.
\newblock \url{http://www.rfpro.com/}.
\newblock (Accessed on 09/22/2020).

\bibitem[Vir()]{VirtualT67:online}
{V}irtual {T}est {D}rive ({VTD}) – {C}omplete {T}ool-{C}hain for {D}riving
  {S}imulation.
\newblock \url{https://www.mscsoftware.com/product/virtual-test-drive}.
\newblock (Accessed on 09/22/2020).

\bibitem[Brockman et~al.(2016)Brockman, Cheung, Pettersson, Schneider,
  Schulman, Tang, and Zaremba]{brockman2016openai}
G.~Brockman, V.~Cheung, L.~Pettersson, J.~Schneider, J.~Schulman, J.~Tang, and
  W.~Zaremba.
\newblock {OpenAI Gym}.
\newblock \emph{arXiv preprint arXiv:1606.01540}, 2016.

\bibitem[Chevalier-Boisvert et~al.(2018)Chevalier-Boisvert, Golemo, Cao, Mehta,
  and Paull]{duckieto16:online}
M.~Chevalier-Boisvert, F.~Golemo, Y.~Cao, B.~Mehta, and L.~Paull.
\newblock Duckietown environments for openai gym.
\newblock \url{https://github.com/duckietown/gym-duckietown}, 2018.

\bibitem[Naoto et~al.(2020)Naoto, Alexander, Billy, and
  Kumar]{ugonamak4:online}
Y.~Naoto, K.~Alexander, Z.~Billy, and A.~Kumar.
\newblock Gym\_torcs.
\newblock \url{https://github.com/ugo-nama-kun/gym_torcs}, 2020.

\bibitem[Palanisamy(2019)]{palanisamy2019multiagent}
P.~Palanisamy.
\newblock Multi-agent connected autonomous driving using deep reinforcement
  learning, 2019.

\bibitem[Anirban et~al.()Anirban, Sohan, Buridi, and Meha]{madrassi35:online}
S.~Anirban, R.~Sohan, A.~Buridi, and K.~Meha.
\newblock {MADR}a{S}: Multi-agent driving simulator.
\newblock \url{https://github.com/madras-simulator/MADRaS}.
\newblock (Accessed on 09/22/2020).

\bibitem[Atlantic()]{InsideWa70}
T.~Atlantic.
\newblock Inside {Waymo}'s secret world for training self-driving cars.
\newblock
  \url{https://www.theatlantic.com/technology/archive/2017/08/inside-waymos-secret-testing-and-simulation-facilities/537648/}.
\newblock (Accessed on 09/22/2020).

\bibitem[Vogt()]{TheDisen76}
K.~Vogt.
\newblock The disengagement myth.
\newblock \url{https://medium.com/cruise/the-disengagement-myth-1b5cbdf8e239}.
\newblock (Accessed on 09/22/2020).

\bibitem[Anguelov()]{DragoAng38}
D.~Anguelov.
\newblock Presentation at {MIT} course on self-driving cars.
\newblock
  \url{https://www.youtube.com/watch?v=Q0nGo2-y0xY&feature=youtu.be&t=1920&ab_channel=LexFridman}.
\newblock (Accessed on 09/22/2020).

\bibitem[Peng et~al.(2017)Peng, Wen, Yang, Yuan, Tang, Long, and
  Wang]{peng2017multiagent}
P.~Peng, Y.~Wen, Y.~Yang, Q.~Yuan, Z.~Tang, H.~Long, and J.~Wang.
\newblock Multiagent bidirectionally-coordinated nets: Emergence of human-level
  coordination in learning to play starcraft combat games.
\newblock \emph{arXiv preprint arXiv:1703.10069}, 2017.

\bibitem[Tan(1993)]{tan1993multi}
M.~Tan.
\newblock Multi-agent reinforcement learning: Independent vs. cooperative
  agents.
\newblock In \emph{Proceedings of the tenth international conference on machine
  learning}, pages 330--337, 1993.

\bibitem[Sutton et~al.(1999)Sutton, McAllester, Singh, Mansour,
  et~al.]{sutton1999policy}
R.~S. Sutton, D.~A. McAllester, S.~P. Singh, Y.~Mansour, et~al.
\newblock Policy gradient methods for reinforcement learning with function
  approximation.
\newblock In \emph{NIPS}, volume~99, pages 1057--1063, 1999.

\bibitem[Mnih et~al.(2016)Mnih, Badia, Mirza, Graves, Lillicrap, Harley,
  Silver, and Kavukcuoglu]{mnih2016asynchronous}
V.~Mnih, A.~P. Badia, M.~Mirza, A.~Graves, T.~Lillicrap, T.~Harley, D.~Silver,
  and K.~Kavukcuoglu.
\newblock Asynchronous methods for deep reinforcement learning.
\newblock In \emph{International Conference on Machine Learning}, pages
  1928--1937, 2016.

\bibitem[Wen et~al.(2018)Wen, Yang, Luo, Wang, and Pan]{wen2018probabilistic}
Y.~Wen, Y.~Yang, R.~Luo, J.~Wang, and W.~Pan.
\newblock Probabilistic recursive reasoning for multi-agent reinforcement
  learning.
\newblock In \emph{International Conference on Learning Representations}, 2018.

\bibitem[Tian et~al.(2019)Tian, Wen, Gong, Punakkath, Zou, and
  Wang]{tian2019regularized}
Z.~Tian, Y.~Wen, Z.~Gong, F.~Punakkath, S.~Zou, and J.~Wang.
\newblock A regularized opponent model with maximum entropy objective.
\newblock \emph{arXiv preprint arXiv:1905.08087}, 2019.

\bibitem[Foerster et~al.(2018)Foerster, Farquhar, Afouras, Nardelli, and
  Whiteson]{comafoerster}
J.~N. Foerster, G.~Farquhar, T.~Afouras, N.~Nardelli, and S.~Whiteson.
\newblock Counterfactual multi-agent policy gradients.
\newblock In S.~A. McIlraith and K.~Q. Weinberger, editors, \emph{Proceedings
  of the Thirty-Second {AAAI} Conference on Artificial Intelligence, New
  Orleans, Louisiana, USA, February 2-7, 2018}. {AAAI} Press, 2018.

\bibitem[Sunehag et~al.(2018)Sunehag, Lever, Gruslys, Czarnecki, Zambaldi,
  Jaderberg, Lanctot, Sonnerat, Leibo, Tuyls, et~al.]{sunehag2018value}
P.~Sunehag, G.~Lever, A.~Gruslys, W.~M. Czarnecki, V.~F. Zambaldi,
  M.~Jaderberg, M.~Lanctot, N.~Sonnerat, J.~Z. Leibo, K.~Tuyls, et~al.
\newblock Value-decomposition networks for cooperative multi-agent learning
  based on team reward.
\newblock In \emph{AAMAS}, pages 2085--2087, 2018.

\bibitem[Rashid et~al.(2018)Rashid, Samvelyan, Schroeder, Farquhar, Foerster,
  and Whiteson]{rashid2018qmix}
T.~Rashid, M.~Samvelyan, C.~Schroeder, G.~Farquhar, J.~Foerster, and
  S.~Whiteson.
\newblock Qmix: Monotonic value function factorisation for deep multi-agent
  reinforcement learning.
\newblock In \emph{International Conference on Machine Learning}, pages
  4295--4304, 2018.

\bibitem[Son et~al.(2019)Son, Kim, Kang, Hostallero, and Yi]{son2019qtran}
K.~Son, D.~Kim, W.~J. Kang, D.~E. Hostallero, and Y.~Yi.
\newblock Qtran: Learning to factorize with transformation for cooperative
  multi-agent reinforcement learning.
\newblock In \emph{International Conference on Machine Learning}, pages
  5887--5896, 2019.

\bibitem[Mahajan et~al.(2019)Mahajan, Rashid, Samvelyan, and
  Whiteson]{mahajan2019maven}
A.~Mahajan, T.~Rashid, M.~Samvelyan, and S.~Whiteson.
\newblock Maven: Multi-agent variational exploration.
\newblock In \emph{Advances in Neural Information Processing Systems}, pages
  7613--7624, 2019.

\bibitem[Osogami and Raymond(2019)]{osogami2019determinantal}
T.~Osogami and R.~Raymond.
\newblock Determinantal reinforcement learning.
\newblock In \emph{Proceedings of the AAAI Conference on Artificial
  Intelligence}, volume~33, pages 4659--4666, 2019.

\bibitem[Zhang et~al.(2018)Zhang, Yang, Liu, Zhang, and
  Ba{\c{s}}ar]{zhang2018finite}
K.~Zhang, Z.~Yang, H.~Liu, T.~Zhang, and T.~Ba{\c{s}}ar.
\newblock Finite-sample analysis for decentralized batch multi-agent
  reinforcement learning with networked agents.
\newblock \emph{arXiv preprint arXiv:1812.02783}, 2018.

\end{thebibliography}
